\documentclass[12pt]{article}

\usepackage{np}
\usepackage{times}

\usepackage{graphicx}
\usepackage{caption}
\usepackage{hyperref}
\usepackage{color}
\usepackage{bm}
\usepackage{lineno}

\title{Beating the Standard Quantum Limit under Ambient Conditions with Solid-State Spins}

\author
{Tianyu Xie$^{1,2,3\dag}$, Zhiyuan Zhao$^{1,2,3\dag}$, Xi Kong$^{4}$, Wenchao Ma$^{1,2,3\ddag}$\\
Mengqi Wang$^{1,2,3}$, Xiangyu Ye$^{2}$, Pei Yu$^{2}$, Zhiping Yang$^{2}$, Shaoyi Xu$^{2}$\\
Pengfei Wang$^{1,2,3}$, Ya Wang$^{2,3}$, Fazhan Shi$^{1,2,3\ast}$ $\&$ Jiangfeng Du$^{1,2,3\ast}$\\
\\
\normalsize{$^{1}$Hefei National Laboratory for Physical Sciences at the Microscale,}\\
    \normalsize{University of Science and Technology of China, Hefei 230026, China}\\
\normalsize{$^{2}$CAS Key Laboratory of Microscale Magnetic Resonance and Department of Modern Physics,}\\
	\normalsize{University of Science and Technology of China, Hefei 230026, China}\\
\normalsize{$^{3}$Synergetic Innovation Center of Quantum Information and Quantum Physics,}\\
	\normalsize{University of Science and Technology of China, Hefei 230026, China}\\
\normalsize{$^{4}$National Laboratory of Solid State Microstructures and Department of Physics,}\\
    \normalsize{Nanjing University, Nanjing 210093, China}\\
\\

\normalsize{$^\dag$These authors contributed equally to this work.}\\
\normalsize{$\ddag$Present address: Department of Chemistry, Massachusetts Institute of Technology,}\\
    \normalsize{Cambridge, Massachusetts 02139, USA}\\
\normalsize{$^\ast$E-mail: fzshi@ustc.edu.cn}\\
\normalsize{$^\ast$E-mail: djf@ustc.edu.cn}\\
}

\date{}

\begin{document}

\captionsetup[figure]{labelfont={bf},labelsep=period}
\baselineskip24pt

\maketitle

\textbf{Precision measurement plays a crucial role in all fields of science. The use of entangled sensors in quantum metrology improves the precision limit from the standard quantum limit (SQL) to the Heisenberg limit (HL)~\cite{beating_SQL_2004, Quantum_metrology_2018}. To date, most experiments beating the SQL are performed on the sensors which are well isolated under extreme conditions~\cite{penning_2016,14_qubits_entanglement_2011,retrodiction_2020,100_times_2016, Deterministic_2016,Entanglement_clock_2020,twin_Fock_2017, Squeezed_clock_2016,LIGO_2013,four_photons_2007,mechanics_2019}. However, it has not been realized in solid-state spin systems at ambient conditions, owing to its intrinsic complexity for the preparation and survival of pure and entangled quantum states. Here we show a full interferometer sequence beating the SQL by employing a hybrid multi-spin system, namely the nitrogen-vacancy (NV) defect in diamond. The interferometer sequence starts from a deterministic and joint initialization, undergoes entanglement and disentanglement of multiple spins, and ends up with projective measurement. In particular, the deterministic and joint initialization of NV negative state, NV electron spin, and two nuclear spins is realized at room temperature for the first time. By means of optimal control, non-local gates are implemented with an estimated fidelity above the threshold for fault-tolerant quantum computation. With these techniques combined, we achieve two-spin interference with a phase sensitivity of 1.79 ± 0.06 dB beyond the SQL and three-spin 2.77 ± 0.10 dB. Moreover, the deviations from the HL induced by experimental imperfections are completely accountable. The techniques used here are of fundamental importance for quantum sensing and computing~\cite{sensing_2017, error_correction_2014}, and naturally applicable to other solid-state spin systems~\cite{silicon_2013, dot_2020, erbium_2013, silicon_carbide_2020}.}

Over the last decades, much effort and progress has been made to realize sub-SQL measurements in many systems such as trapped ions~\cite{penning_2016, 14_qubits_entanglement_2011}, cold thermal atoms~\cite{retrodiction_2020, 100_times_2016, Deterministic_2016}, Bose-Einstein condensates (BEC)~\cite{Entanglement_clock_2020, twin_Fock_2017, Squeezed_clock_2016}, photons~\cite{LIGO_2013, four_photons_2007}, and mechanical systems~\cite{mechanics_2019}. These systems are appropriate for some extraordinary applications including precision tests of the fundamental laws of physics~\cite{new_physics_2018}, atomic clocks~\cite{Entanglement_clock_2020, Squeezed_clock_2016} and gravitational wave detection~\cite{LIGO_2013}. However, their sensors only work well under extreme conditions, and thus not suitable for most realistic measurements demanding moderate conditions or a high spatial resolution, e.g., microscale imaging at ambient conditions. Fortunately, a recently emergent solid-state spin system,  nitrogen-vacancy centers in diamond, provides such an approach. With this atomic-scale sensor, electron and nuclear magnetic resonance of single molecules have been accomplished~\cite{ESR_2015, NMR_2016}, while nanoscale magnetic imaging combined with an atomic force microscope is being put into practice~\cite{periodic_2017, hopping_2014}. Though there are some NV-based work related to the SQL~\cite{Fourier_2020, dynamic_2012, enhanced_2015}, the canonical method of using entanglement to beat the SQL remains elusive. The main challenge lies in the difficulty of deterministically initializing solid-state spins to a pure state and deeply entangling the spins at ambient conditions.

\begin{figure*}\center
\includegraphics[width=1.0\textwidth]{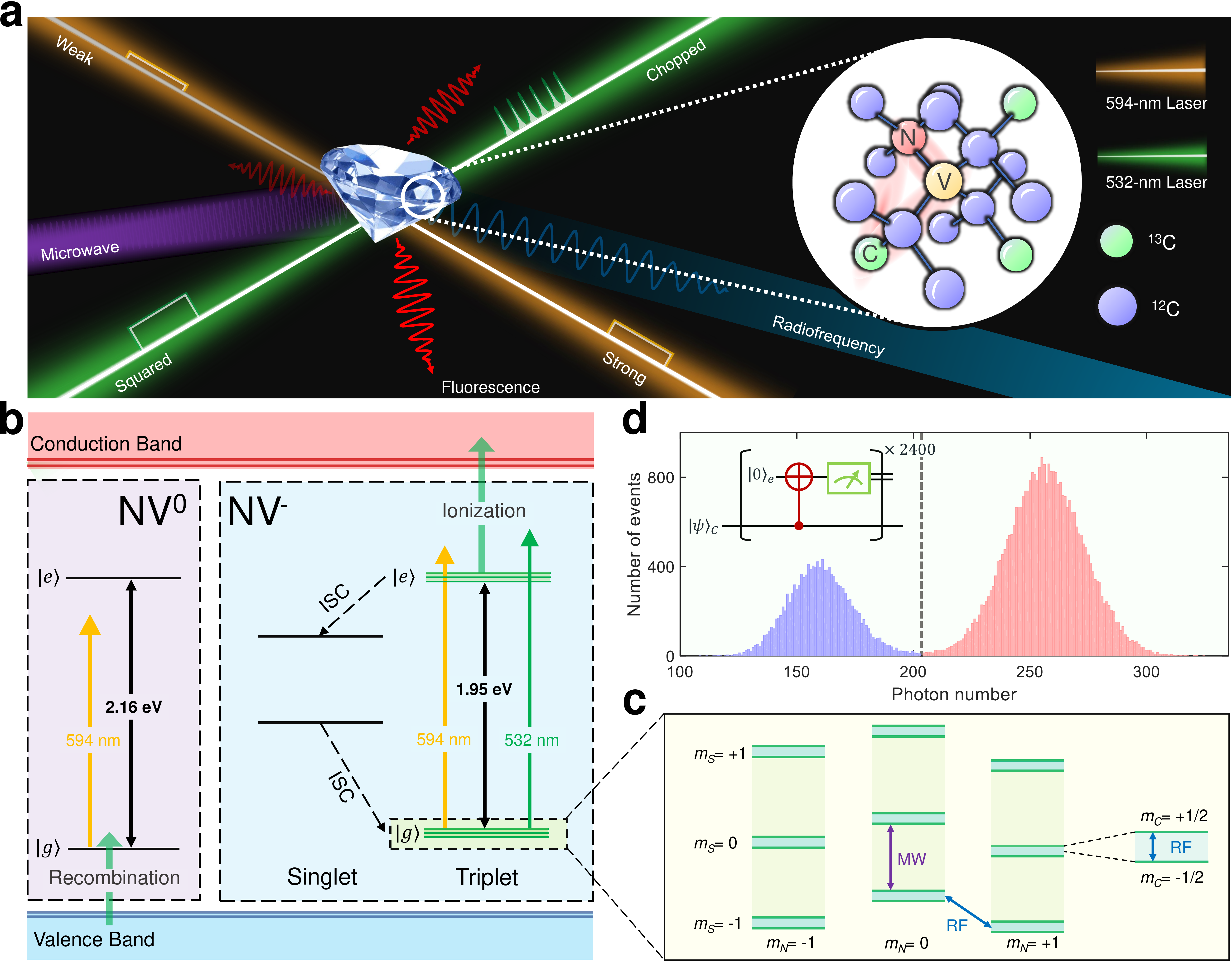}
\caption{\label{overview}\textbf{Dynamics and control of a single NV center.} \textbf{a,} Diagram of a single NV center in diamond. It is driven by various laser beams and coherently controlled by MW and RF pulses. Squared: squared 532-nm green laser for readout of the NV electron spin and mixing charge states; chopped: chopped laser sequence for a better polarization of the electron spin without destroying charge state; strong: strong (4 \textmu W) 594-nm orange laser for readout of charge state in real-time feedback; weak (0.18 \textmu W) orange laser for single-shot readout of charge state. The NV electron spin ($S=1$), the attendant $^{14}$N nuclear spin ($I=1$) and one of randomly distributed $^{13}$C nuclear spins ($I=1/2$) constitute the interferometer. \textbf{b,} Level diagrams of NV negative and neutral states denoted by NV$^{-}$ and NV$^{0}$, and the corresponding dynamics driven by 532-nm green laser and 594-nm orange laser. \textbf{c,} Spin level structure of the ground state of NV$^{-}$ triplet. MW and RF pulses are used to coherently manipulate the NV electron spin and two nuclear spins ($^{14}$N and $^{13}$C). \textbf{d,} Projective measurement of the $^{13}$C nuclear spin. The dashed line denotes the threshold to determine which state it stays in.
}
\end{figure*}

Here we demonstrate the entire measurement process for an entangled interferometer based on a single NV defect center. The NV center in diamond is formed by a substitutional nitrogen atom and an adjacent vacancy as shown in Fig. \ref{overview}a. The NV electron spin ($S=1$), the attendant $^{14}$N nuclear spin ($I=1$) and one of randomly distributed $^{13}$C nuclear spins ($I=1/2$) constitute the interferometer. Through hyperfine interactions between the electron spin and nuclear spins, multi-qubit quantum entanglement is producible. Complicated dynamics involving multiple degrees of freedom (charges, orbits and spin angular momenta) are driven by 532-nm green laser and 594-nm orange laser for the initialization and readout of the NV charge state and the electron spin detailed in Fig. \ref{overview}b. Microwave (MW) and radio-frequency (RF) pulses imposed by a coplanar waveguide can coherently manipulate the electron spin and two nuclear spins both locally and non-locally (Fig. \ref{overview}c).

Projective measurement is an indispensable ingredient to beat the SQL. It is implemented via the sequence of the inset of Fig. \ref{overview}d for a nuclear spin~\cite{singleshot_2010,C_singleshot_2013}. The sequence eliminates the classical noise via thousands of readouts of the nuclear spin whose information is correlated with the electron spin via a $\rm C_{n}NOT_{e}$ (an electron NOT gate conditional on nuclear state) before the electron spin readout during every cycle. The non-destructive nature of the nuclear spin after every readout is brought about by a high magnetic field (8066 G in our setup). The result of implementing the sequence is a two-peak statistic of photon counts for the $^{13}$C nuclear spin (Fig. \ref{overview}d). We obtain a readout fidelity of 99.14(4)$\%$ by setting an optimal threshold to determine the spin state.

\begin{figure*}\center
\includegraphics[width=1.0\textwidth]{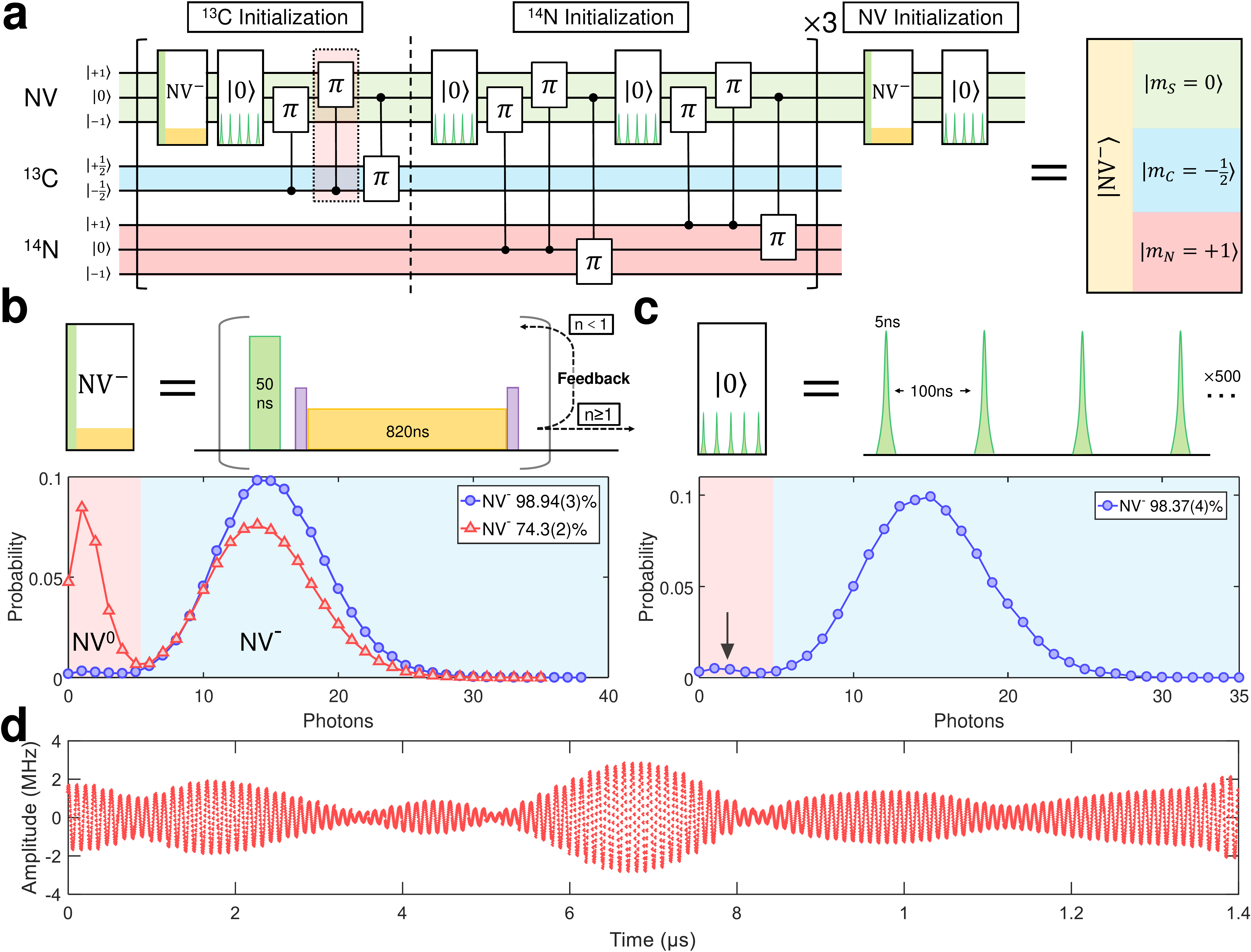}
\caption{\label{initialization}\textbf{Deterministic and joint initialization.} \textbf{a,} The overall sequence for initialization of the joint state $|{\rm NV}^{-}, m_{S}=0\rangle \otimes |m_{C}=-1/2\rangle \otimes |m_{N}=+1\rangle$. \textbf{b,} NV$^{-}$ preparation by real-time feedback. Upper panel: the implemented sequence, photon counting during the interval between two purple boxes; lower panel: the results of single-shot readout of charge state with (blue circles, 98.94$\%$ NV$^{-}$) and without (red triangles, 74.3$\%$ NV$^{-}$) NV$^{-}$ preparation. \textbf{c,} The $|m_{S}=0\rangle$ preparation of the NV electron spin. Upper panel: chopped laser sequence for a better polarization to the $|m_{S}=0\rangle$ state; lower panel: the survival probability of the NV negative state under chopped laser sequence (98.37$\%$/98.94$\%$$\approx$99.42$\%$) measured by the single-shot readout of charge state. The slight bulge indicated by the black arrow represents the damage 0.58$\%$ caused by chopped laser sequence. \textbf{d,} An example of shaped pulse sequence with optimal control (see Supplementary Section 2.2). It is used in polarization transfer from the electron spin to two nuclear spins and non-local gates in multi-qubit interference in Fig. \ref{interference}a.
}
\end{figure*}

Deterministic initialization to a pure enough state is equally crucial. Especially, it is rather difficult for solid-state spin system at room temperature resulting from the complicated dynamics of the local defect electron (Fig. \ref{overview}b) exposed to a thermal phonon sea and energy bands, unlike well isolated ions trapped in electromagnetic field~\cite{penning_2016, 14_qubits_entanglement_2011}. Existing methods can probabilistically initialize NV negative state and nuclear spins heralded by measurement outcomes with abandoning a considerable proportion of data, which will dramatically reduce the sensitivity~\cite{error_correction_2014}. Here we propose a brand new scheme to deterministically and jointly initialize NV charge state, electron spin and two nuclear spins with a high fidelity, in combination with two novel techniques developed recently~\cite{charge_2020, polarization_2020}. The overall sequence, depicted in Fig. \ref{initialization}a, initializes our system to the joint state $|{\rm NV}^{-}, m_{S}=0\rangle \otimes |m_{C}=-1/2\rangle \otimes |m_{N}=+1\rangle$ with a fidelity of 93.3(3)$\%$ (see Supplementary Section 4.4).

The scheme includes three elaborately devised techniques, i.e., real-time feedback for NV negative state preparation (Fig. \ref{initialization}b)~\cite{charge_2020}, chopped laser sequence for a better spin polarization ~\cite{polarization_2020,polarization_2019} without destroying charge state (Fig. \ref{initialization}c) and exquisite optimal control and population shelving for polarization transfer to nuclear spins (Fig. \ref{initialization}d). As shown in Fig. \ref{overview}b, 594-nm orange laser excites the NV negative state but does not excite the neutral state. Therefore, the negative state is prepared if one photon is collected during laser illumination, and if not do it again. Before photon counting during every cycle, the green laser is switched on to mix charge states (Fig. \ref{initialization}b). The fidelity is estimated to be 98.94(3)$\%$ by single-shot readout of charge state (see Supplementary Section 4.1). Conventionally, a squared pulse of green laser is applied for polarizing the electron spin to $|m_{S}=0\rangle$ with a fidelity of 90.0(3)$\%$. Here it is replaced by a chopped laser sequence plotted in Fig. \ref{initialization}c with an improved fidelity of 97.8(2) $\%$ (see Supplementary Section 4.2). Most strikingly, the NV negative state can survive through it with a near-unit probability of 99.42(5)$\%$ obtained from the lower part of Fig. \ref{initialization}c. With these two techniques combined, the defect is initialized in $|{\rm NV}^{-}, m_{S}=0\rangle$. After that, polarization transfer from the electron spin via swap-like gates initializes the nuclear spins with the estimated fidelities of 98.34(13)$\%$ for $^{13}$C and 98.71(18)$\%$ for $^{14}$N (see Supplementary Section 4.3). In order to reduce the polarization loss induced by imperfection of non-local operation, the technique of optimal control is harnessed to achieve robust and high-fidelity state manipulation with one sequence displayed in Fig. \ref{initialization}d. To further diminish the residual error, the population of the state flipped by $\rm C_{n}NOT_{e}$ is depleted by shelving to $|m_{S}=+1\rangle$ indicated by the dashed box of Fig. \ref{initialization}a. The techniques aforementioned can be compatibly assembled, mainly benefitting from two vast gaps: one between the green laser duration $\sim$4 \textmu s in real-time feedback and the lifetimes of nuclear spins under laser illumination $\sim$40 ms for $^{13}$C and $\sim$10 ms for $^{14}$N; the other between the laser duration $\sim$2.5 \textmu s in chopped laser sequence and the lifetime of NV negative state $\sim$0.4 ms illuminated by chopped laser sequence.

\begin{figure*}\center
\includegraphics[width=1.0\textwidth]{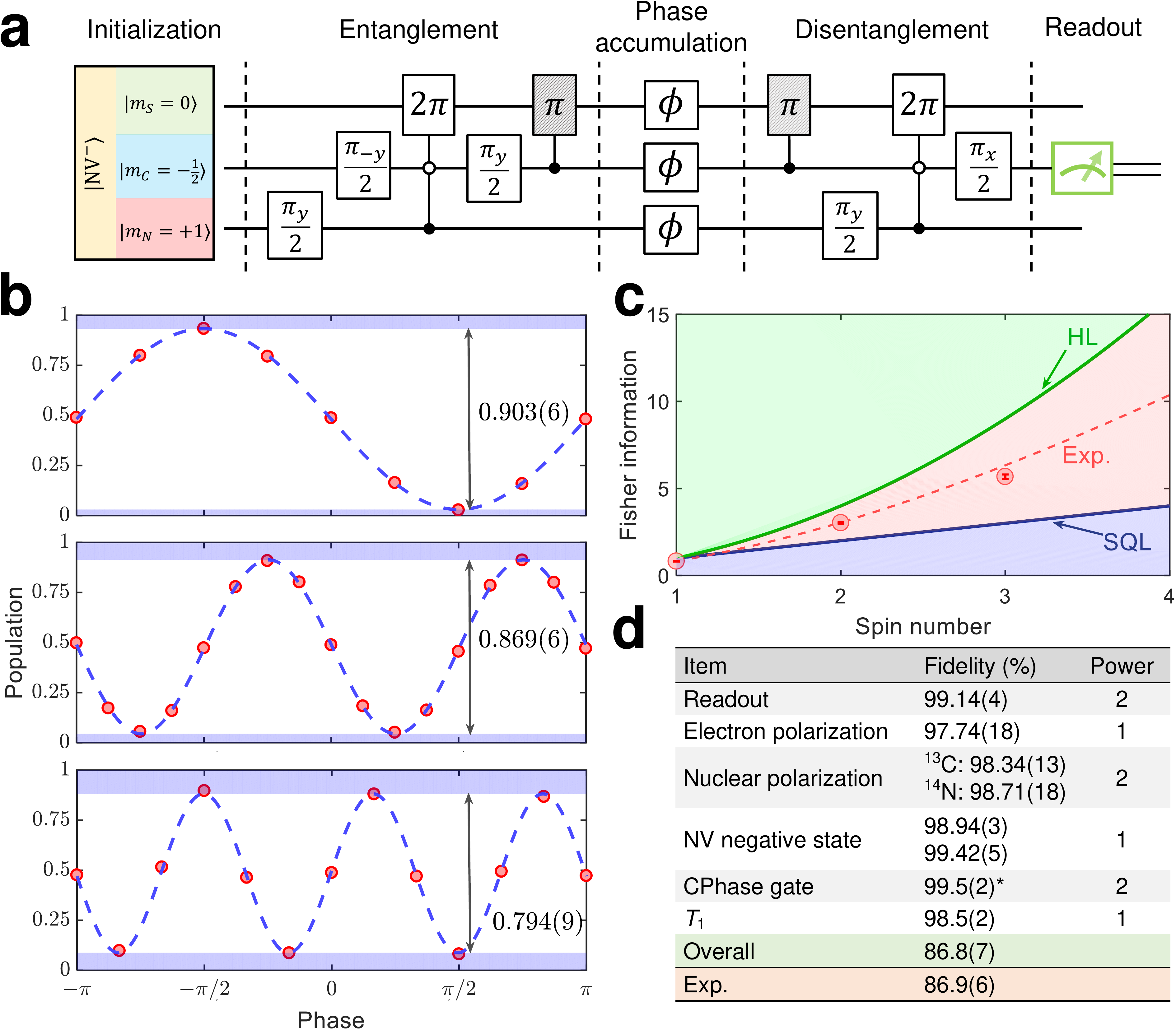}
\caption{\label{interference}\textbf{Multi-qubit entangled interferometer.} \textbf{a,} Quantum circuit for three-spin interference. Excluding two $\rm C_{n}NOT_{e}$ gates (the shadowed boxes) gives two-spin interference.  It consists of five parts separated by dashed lines; that is, initialization, entanglement, phase accumulation, disentanglement and readout. \textbf{b,} Interference patterns for one ($^{13}$C), two ($^{13}$C and $^{14}$N) and three spins from top to bottom. The interference visibility is indicated by two-head arrows. \textbf{c,} Quantum Fisher information of our interferometer as a function of spin number. The function of the dashed line is $N^{2}(0.91\times 0.96^{(N - 1)})^{2}$ considering our experimental imperfections, in comparison with the SQL of $N$ and the HL of $N^{2}$. The error bars are inside the empty circles representing experimental data. \textbf{d,} Table of error budget for two-spin interference. The fidelity of every item is measured by an independent experiment except the item labeled by an asterisk ($\ast$) with the fidelity estimated by simulation. The overall fidelity equals $\prod_{i}{(\rm{Fidelity}_{i})^{\rm{Power}_{i}}}$. The fidelity of the NV negative state comes from two parts, the preparation fidelity by real-time feedback 98.94(3)$\%$ and the survival probability 99.42(5)$\%$ under the chopped laser sequence. All errors in parentheses stand for 1 standard deviation.
}
\end{figure*}

The main sequence performing the entanglement-based interference is displayed in Fig. \ref{interference}a. Here we prepared the Greenberger-Horne-Zeilinger (GHZ) state that is maximally metrologically useful. In comparison to directly measuring the parity of the GHZ state after acquiring an overall phase $N\phi$~\cite{14_qubits_entanglement_2011}, we disentangle the state and the phase $N\phi$ enters into only one of the spins with other spins separable. In this way, the projective measurement of only one spin is needed, leading to the reduction of the readout error (the readout fidelity for $^{14}$N is only 97.02(6)$\%$). To entangle nuclear spins, we adopted a novel scheme used in~\cite{error_correction_2014} to realize a conditional $\pi$-phase (CPhase) gate on the electron spin. Furthermore, the technique of optimal control is merged to achieve high-fidelity control. The shaped pulse of 99.5$\%$ fidelity is exquisitely optimized by the GRAPE algorithm (gradient ascent pulse engineering) with a similar looking of Fig. \ref{initialization}d (see Supplementary Section 2.2, 5.1). For three-spin entanglement, an extra $\rm C_{n}NOT_{e}$ is implemented to entangle the NV electron spin. Two-spin (two nuclear spins) and three-spin interference patterns are given by Fig. \ref{interference}b (one-spin interference as reference) and the interference visibility is indicated by a two-head arrow. The quantum Fisher information is calculated from the visibility accordingly, and plotted as a function of spin number presented in Fig. \ref{interference}c (see Supplementary Section 1.3). It is clearly exhibited that Fisher information of our entangled interferometer is well beyond the SQL (two-spin 1.79 ± 0.06 dB and three-spin 2.77 ± 0.10 dB) and gets close to the HL. A scaling of Fisher information $N^{2}(0.91\times 0.96^{(N - 1)})^{2}$ is expected with experimental imperfections considered, and saturates at 24 spins (see Supplementary Section 5.3). Besides, we scrutinize all experimental imperfections listed in the table of Fig. \ref{interference}d. The overall fidelity 86.8(7)$\%$ is in an excellent agreement with the visibility of two-spin interference 86.9(6)$\%$ (see Supplementary Section 5.2).

\begin{figure*}\center
\includegraphics[width=1.0\textwidth]{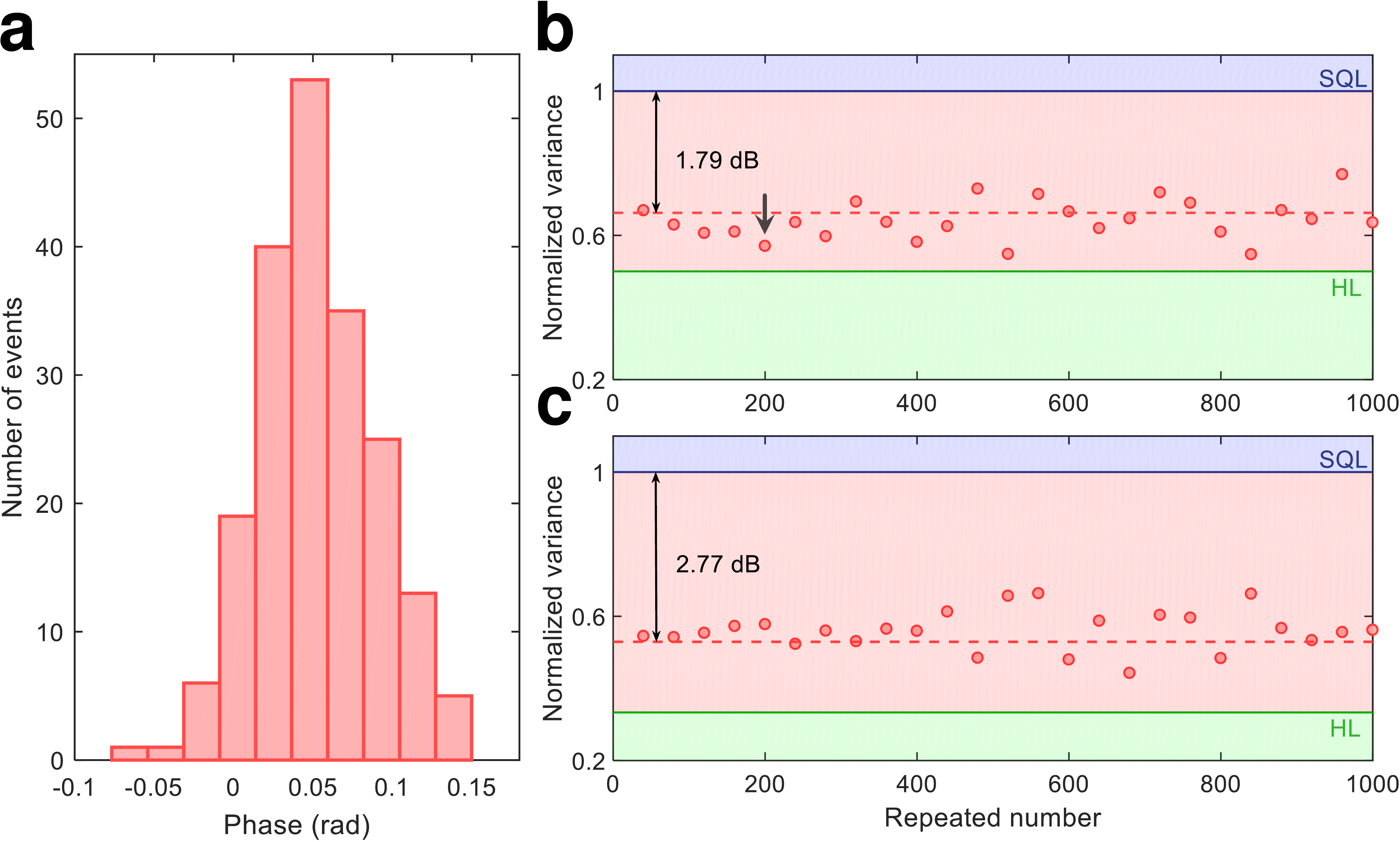}
\caption{\label{variance}\textbf{Phase variance measured under a real noise environment.} \textbf{a,} Histogram of phase estimates. Every phase estimate is obtained by repeating the measurement 200 times. It is used to determine the phase variance in \textbf{b} and \textbf{c}. \textbf{b (c)}, Phase variance of two-spin (three-spin) interference with different repeated number $\nu$ of the measurement. All variances are normalized to those of the SQL ($\frac{1}{2\nu}$ for two spins and $\frac{1}{3\nu}$ for three spins). The variance of the distribution in \textbf{a} is the point indicated by the black arrow in \textbf{b}.
}
\end{figure*}

Finally, the statistics of a measured phase $\phi=\pi/60$ for two-spin interference ($\phi=\pi/90$ three-spin) under a real magnetic noise environment are studied in Fig. \ref{variance}a. The value of every phase in the histogram is obtained by performing independent and identical experiments 200 times. From the histogram, the variance of the measured phase is acquired, which corresponds to the point of Fig. \ref{variance}b indicated by the black arrow. All points in Fig. \ref{variance}b,c spread around the expected line well below the SQL, signifying that our entangled interferometer truly beats the SQL. It is noticed that a magnetic fluctuation of 0.01 G will induce a phase fluctuation of $\sim$0.015, while the uncertainty is 0.018 for the average of 1000 times of our two-spin entangled interferometer. It means the magnetic field stability is well below 0.01 G in two hours for data acquisition, rather good for our 8066 G magnetic field.

In conclusion, the SQL is beaten in solid-state spin system at ambient conditions for the first time. The statistical uncertainty of the estimated phase is 1.79 dB below the SQL for two qubits and 2.77 dB for three qubits. For a sub-SQL interference, we have implemented all elements including deterministic initialization, high-fidelity entanglement and disentanglement, and projective measurement. The joint state $|{\rm NV}^{-}, m_{S}=0\rangle \otimes |m_{C}=-1/2\rangle \otimes |m_{N}=+1\rangle$ is deterministically prepared at room temperature for the first time. Non-local gates are realized by exquisite optimal control with the accuracy required for fault-tolerant quantum computation ~\cite{surface_code_2011}. These basic ingredients have great significance for NV-based quantum sensing and computing~\cite{sensing_2017, error_correction_2014}, and could be extended to other solid-state spin systems~\cite{silicon_2013, dot_2020, erbium_2013, silicon_carbide_2020}. In the future, the technique toolbox adopted here can be directly applied to the circumstance of magnetic resonance spectroscopy and imaging at microscale  with a higher magnetic sensitivity by using multiple NV electron spins. The genuine entanglement among multiple NV spins could be generated either by direct magnetic interaction~\cite{NV_pair_2014} or coupling to a common bus~\cite{coupling_2020, cavity_2015}.

\bibliographystyle{Nature}

\clearpage

\section*{Methods}
\subsection*{Diamond sample}
The targeted NV center resides in a bulk diamond whose top face is perpendicular to the [100] crystal axis and lateral faces are perpendicular to [110]. The nitrogen concentration of the diamond is less than 5 p.p.b. and the abundance of $^{13}$C is at the natural level of 1.1$\%$. The diamond is irradiated using 10-MeV electrons to a total dose of $\sim$1.0 kgy. A solid immersion lens (SIL) is created around the targeted NV center to increase the luminescence rate of the NV to $\sim$400 kcounts/s.
\subsection*{Experimental setup}
The diamond is mounted on a typical optically detected magnetic resonance confocal setup, synchronized with a microwave bridge by a multichannel pulse blaster (Spincore, PBESR-PRO-500). The 532-nm green laser and the 594-nm orange laser for driving NV electron dynamics, and sideband fluorescence (650–800 nm) go through the same oil objective (Olympus, UPLSAPO 100XO, NA 1.40). To protect the NV center’s negative state and longitudinal relaxation time against laser leakage effects, all laser beams pass twice through acousto–optic modulators (AOM) (Gooch $\&$ Housego, power leakage ratio $\sim$1/1,000) before they enter the objective. Chopped laser sequence is realized by feeding a short-pulse sequence into the green-laser AOM. The short-pulse sequence is generated by an arbitrary waveform generator (AWG) (Zurich Instruments, HDAWG4). The fluorescence photons are collected by avalanche photodiodes (APD) (Perkin Elmer, SPCM-AQRH-14) with a counter card (National Instruments, 6612). The ZI AWG also has a built-in counter to perform real-time feedback for preparing NV negative state. The 19.7 GHz and 25.5 GHz microwave pulses for the manipulation of the NV three sublevels are generated from the microwave bridge, coupled with 0.1-10 MHz radio-frequency pulses for the nuclear spins via a diplexer, and fed together into the coplanar waveguide microstructure. The external magnetic field ($\approx8066$ G) is generated from a permanent magnet and aligned parallel to the NV axis through a three-dimensional positioning system. The positioning system, together with the platform holding the diamond and the objective, is placed inside a thermal insulation copper box. The temperature inside the copper box stabilizes down to a sub-mK level through the feedback of the temperature controller (Stanford, PTC10) (see Supplementary Section 3.2).

\section*{Acknowledgements}
We thank Nanyang Xu and Bing Chen (Hefei University of Technology, China) for helpful discussions on the polarization of the NV center by chopped laser sequence.
This work was supported by the National Natural Science Foundation of China (Grant Nos. 91636217, 81788101, 11722544, 11761131011), the National Key R$\&$D Program of China (Grant Nos. 2018YFA0306600 and 2016YFA0502400), the CAS (Grant Nos. GJJSTD20170001 and QYZDY-SSW-SLH004), the Anhui Initiative in Quantum Information Technologies (Grant No. AHY050000), and the Fundamental Research Funds for the Central Universities.

\section*{Author contributions}
J.D. and F.S. supervised the project and proposed the idea. T.X., W.M., F.S. and J.D. designed the experiments. T.X., Z.Z. and X.K. prepared the setup. Z.Z., M.W., P.Y., X.Y., S.X., Z.Y. and Y.W. prepared the diamond sample. T.X. and Z.Z. performed the experiment and the simulation. T.X., Z.Z., F.S. and J.D. wrote the manuscript. All authors analysed the data, discussed the results and commented on the manuscript.

\section*{Competing financial interests}
All authors declare no competing financial interests.

\end{document}

\captionsetup[figure]{labelfont={bf},labelsep=period}
\baselineskip24pt

\maketitle
\pagebreak[4]

\tableofcontents
\pagebreak[4]

\section{A brief introduction to quantum metrology}
\subsection{Fisher information and quantum Fisher information}
Fisher information plays a central role in the theory of parameter estimation. Let us consider a random variable $X$ with a probability distribution $P(x|\theta)$ containing a parameter $\theta$. For an unbiased estimator $\Theta$ for the parameter $\theta$ and $\nu$ independent experiments, the variance of the estimator $\Theta$ gives the precision of the estimation for the parameter $\theta$ , satisfying
\begin{equation}
\label{Rao_bound}
(\Delta\theta)^{2}=\rm{Var}(\Theta) \ge \frac{1}{\nu \it{F(\theta)}}
\end{equation}
with Fisher information
\begin{equation}
\label{Fisher_information}
F(\theta)=\sum_{x}{P(x|\theta)}(\frac{\partial \ln P(x|\theta)}{\partial \theta})^2
\end{equation}
The lower bound given by the right side of Eq. \ref{Rao_bound} is the well-known Cram\'{e}r-Rao bound~\cite{Quantum_metrology_2018}. That is, Fisher information gives the best precision of the estimation for the parameter $\theta$.

Quantum mechanics is intrinsically probabilistic, and thus any measurement is bound to be a parameter estimation. However, the measurements with different bases ${\hat{E}}$ for the same quantum state $\hat{\rho}_{\theta}$ with a phase $\theta$ generate different probability distributions and bring about extra complexity in comparison with classical measurements. Therefore, the maximum of Eq. \ref{Fisher_information} over all possible measurements allowed by quantum mechanics is defined as the quantum Fisher information
\begin{equation}
\label{quantum_Fisher_information}
F_Q[\hat{\rho}_{\theta}] = \max_{\hat{E}}F(\theta) \ge F(\theta)
\end{equation}
which is independent of a specific measurement and the nature of the quantum state $\hat{\rho}_{\theta}$. Correspondingly, the lower bound of the measurement precision is called the quantum Cram\'{e}r-Rao bound
\begin{equation}
\label{quantum_Rao_bound}
(\Delta\theta)^{2} \ge \frac{1}{\nu F_Q[\hat{\rho}_{\theta}]}
\end{equation}
For a pure state $\hat{\rho}_{\theta}=|\Psi_\theta\rangle\langle\Psi_\theta|$, the quantum Fisher information simply reads
\begin{equation}
\label{pure_QFI}
F_Q[|\Psi_\theta\rangle]=\rm{4}(\langle\partial_\theta\Psi_\theta | \partial_\theta\Psi_\theta\rangle-|\langle\partial_\theta\Psi_\theta | \Psi_\theta\rangle|^{\rm{2}})
\end{equation}
with $\partial_\theta\Psi_\theta=\partial\Psi_\theta/\partial\theta$. For most circumstances in quantum metrology, the phase $\theta$ for parameter estimation is encoded by a local Hamiltonian $\hat{H}=\sum_{i=1}^{N}\hat{h}_{i}$ of $N$ particles. Here we only consider the case of two-level system (qubit) where $\hat{h}_i=\hat{\sigma}_{\boldsymbol{n}}^{i}/2$ with $\sigma_{\boldsymbol{n}}^{i}$ the Pauli operator of the $i$th particle. Therefore, with the total angular momentum along the $\boldsymbol{n}$ direction $\hat{J}_{\boldsymbol{n}}=\sum_{i=1}^{N}\hat{\sigma}_{\boldsymbol{n}}^{i}/2 $, the evolution for phase encoding is given by $U(\theta)=\exp(-i\theta\hat{J}_{\boldsymbol{n}})$.
For a pure state $\hat{\rho}_{0}=|\Psi_0\rangle\langle\Psi_0|$ under the evolution, the Eq. \ref{pure_QFI} becomes
\begin{equation}
\label{pure_QFI_J}
F_Q[|\Psi_{\rm{0}}\rangle,U(\theta)]=\rm{4}(\Delta\it{\hat{J}}_{\boldsymbol{n}})^{\rm{2}}
\end{equation}
It is learned from the equation that a pure state with a vast difference of the eigenvalues of $\hat{J}_{\boldsymbol{n}}$ tends to have larger quantum Fisher information.

\subsection{The standard quantum limit, the Heisenberg limit and some important quantum states}
For a separable state of $N$ qubits $\hat{\rho}_{\rm sep}$ under the evolution $U(\theta)$, it is easy to prove that the quantum Fisher information is upper bounded by
\begin{equation}
\label{SQL}
F_Q[\hat{\rho}_{\rm sep},U(\theta)] \le N
\end{equation}
It is the so-called standard quantum limit (SQL). The equality holds when the separable state is coherent spin states (CSS)
\begin{equation}
\label{CSS}
|\alpha, \phi, N\rangle=\otimes_{i=1}^{N}[\cos(\frac{\alpha}{2})|0\rangle_i+e^{i\phi}\sin(\frac{\alpha}{2})|1\rangle_i]
\end{equation}
with the mean spin direction $(\sin(\alpha)cos(\phi),\sin(\alpha)sin(\phi),\cos(\alpha))$ perpendicular to the direction $\boldsymbol{n}$ of the evolution. It can be easily verified by applying Eq. \ref{pure_QFI_J} to the CSS. If Eq. \ref{SQL} is not satisfied,
\begin{equation}
\label{entanglement_criterion}
F_Q[\hat{\rho},U(\theta)] > N
\end{equation}
it is sufficient to say that the state is entangled. Eq. \ref{entanglement_criterion} is a criterion of entanglement and it measures metrologically useful entanglement. For any state $\hat{\rho}$ under the evolution $U(\theta)$, no matter whether it is separable or entangled, the quantum Fisher information has an ultimate limit
\begin{equation}
\label{HL}
F_Q[\hat{\rho},U(\theta)] \le N^2
\end{equation}
which is called the Heisenberg limit (HL). By using Eq. \ref{pure_QFI_J}, we find that the equal superposition of two eigenvectors of $\hat{J}_{\boldsymbol{n}}$ with the maximum and the minimum eigenvalue ($N/2$ and$-N/2$)
\begin{equation}
\label{GHZ}
|\rm{GHZ}\rangle=(|00\cdot\cdot\cdot 0\rangle_{\boldsymbol{n}}+|11\cdot\cdot\cdot 1\rangle_{\boldsymbol{n}})/\sqrt{2}
\end{equation}
make the equality hold. The state is called the Greenberger-Horne-Zeilinger (GHZ) state (the NOON state in the language of boson) which we used in our experiments. Other entangled states include spin squeezed states (SSS) squeezed along the direction $\boldsymbol{n}$ with $\xi_R^2<1$~\cite{spin_squeezing}, Dicke states $|m_{\boldsymbol{n}}\rangle$ (eigenvectors of $\hat{J}_{\boldsymbol{n}}$)~\cite{Dicke_states} and the twin-Fock state $|0_{\boldsymbol{n}}\rangle$ (the Dicke state with $m_{\boldsymbol{n}}=0$)~\cite{twin_Fock_2017} with quantum Fisher information beyond the SQL
\begin{equation}
\label{SSS}
F_Q[|SSS\rangle,U(\theta,\boldsymbol{n}_\bot)] \ge N/\xi_R^2
\end{equation}
\begin{equation}
\label{Dicke}
F_Q[|m_{\boldsymbol{n}}\rangle,U(\theta,\boldsymbol{n}_\bot)] = \frac{N^2}{2}-2m^2+N
\end{equation}
\begin{equation}
\label{Twin}
F_Q[|0_{\boldsymbol{n}}\rangle,U(\theta,\boldsymbol{n}_\bot)] = \frac{N^2}{2}+N
\end{equation}
with $U(\theta,\boldsymbol{n}_\bot)=\exp(-i\theta\hat{J}_{\boldsymbol{n}_\bot})$. By the way, the twin-Fock state is also the squeezing limit of the SSS.

\subsection{Quantum Fisher information acquired from interference experiments}
\label{QFI_exp}
For a realistic experiment of preparing the GHZ state, the purity will be reduced by various experimental imperfections, deviating from the HL. The quantum Fisher information of our entangled interferometer is estimated by the method of moments~\cite{Quantum_metrology_2018}
\begin{equation}
\label{real_GHZ}
F_Q[\rho_{\rm{exp}}] \approx \frac{|d\mu/d\theta|^2}{\nu(\Delta\mu)^2}=(\rm{Vis})^2*\it{N}^{\rm{2}}
\end{equation}
where $\mu$ is the parameter of the population we measured in our experiments, $\nu$ is the number of independent and identical experiments and `Vis' stands for the visibility (the amplitude of interference pattern). In the central limit $\nu\longrightarrow\infty$, Eq. \ref{real_GHZ} holds with a high enough precision.

\pagebreak[4]

\section{The system Hamiltonian of solid-state spins and coherent control}
\subsection{The system Hamiltonian}
The NV electron spin ($S=1$) in the ground state of the NV$^-$ triplet, the adjacent $^{14}$N nuclear spin ($I=1$), and one of randomly distributed $^{13}$C nuclear spins ($I=1/2$) comprise a quantum system of multiple spins at nanoscale. The Hamiltonian with an external magnetic field  $B_0\approx$ 8066 G applied along the axis of the NV is given by
\begin{equation}
\label{Hamiltonian_0}
H_0 \approx \overbrace{D S_z^2+\gamma _e B_0 S_z}^{\rm{NV}}+\overbrace{Q (I_z^N)^2-\gamma _N B_0 I_z^N+A_\parallel S_z I_z^N}^{^{14}\rm{N}} + \overbrace{-\gamma _C B_0 I_z^C+A_{zz} S_z I_z^C}^{^{13}\rm{C}}
\end{equation}
where $\gamma _e$, $\gamma _N$ and $\gamma _C$ are the gyromagnetic ratios of the electron spin, the $^{14}$N nuclear spin and the $^{13}$C nuclear spin respectively. $S_z$, $I_z^N$ and $I_z^C$ are the components of three spin operators along the axis of the NV. The hyperfine interactions $A_\parallel$ and $A_{zz}$ are 2164.9 kHz and 375.4 kHz for $^{14}$N and $^{13}$C. $D=2869.73$ MHz and $Q=4945.8$ kHz are the zero-field splittings of the electron spin and the $^{14}$N nuclear spin. MW and RF with multiple frequencies are imposed to coherently control the electron spin and the nuclear spins with the control Hamiltonian
\begin{equation}
\label{Hamiltonian_c}
H_c(t) = \sum_{i}{2(\Omega_i^{MW}(t)\cos(\omega _i^{MW} t+\phi_i^{MW}(t))+\Omega _i^{RF}(t)\cos(\omega _i^{RF} t+\phi_i^{RF}(t))})(S_x+I_x^N+I_x^C)
\end{equation}
We implemented the quantum circuits in the text by transforming into a suitable interaction picture with some rotation-wave approximations (see Section \ref{local_heating}).

\subsection{Non-local gates and optimal control}
\label{local_heating}
\begin{figure*}\center
\includegraphics[width=1.0\textwidth]{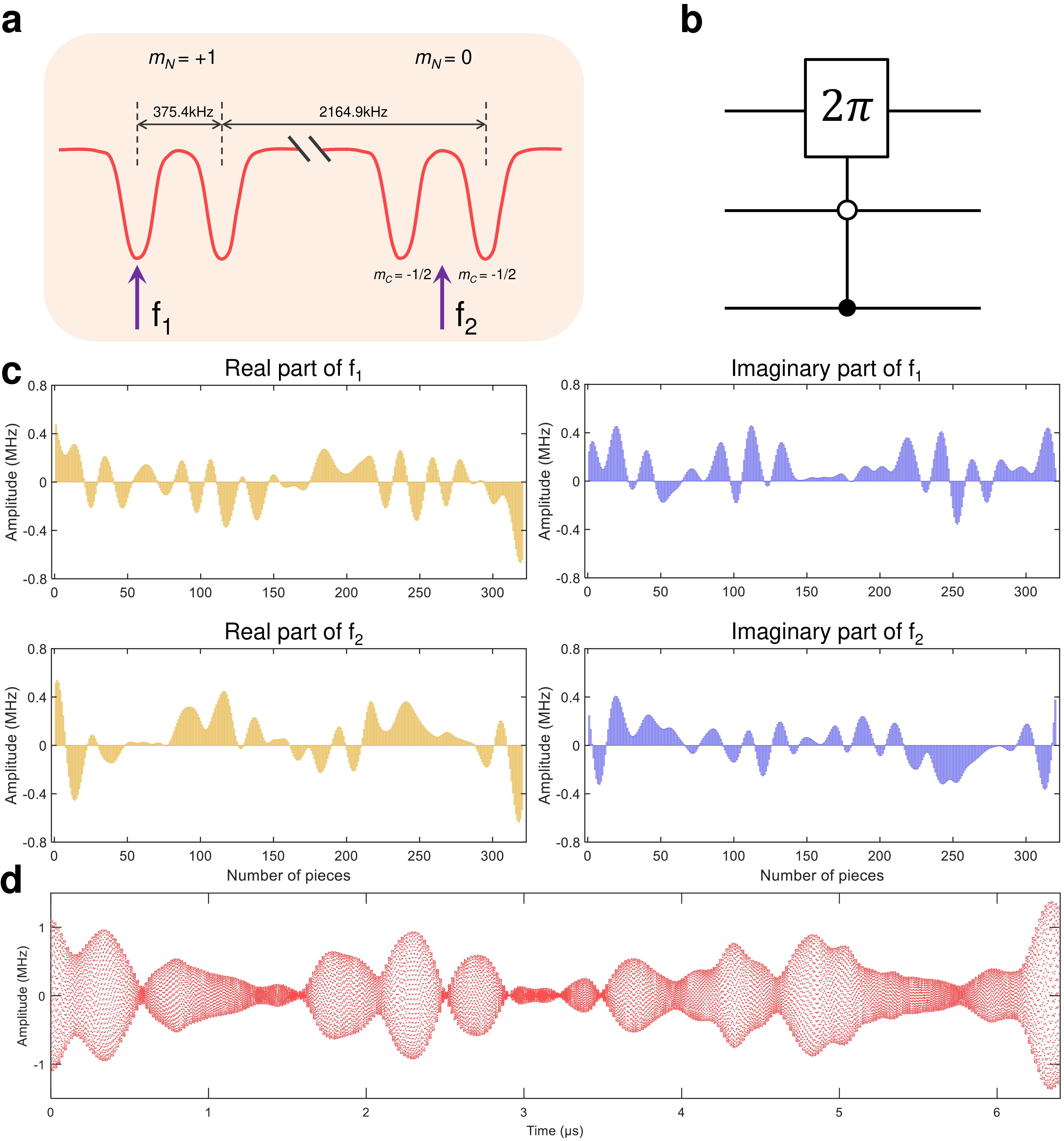}
\caption{\label{grape}\textbf{Optimal control.} \textbf{a,} Two MW frequencies $f_1$ and $f_2$ applied to the hyperfine splittings with an asymmetry. \textbf{b,} The CPhase gate realized by optimal control. \textbf{c,} The pulse sequence of the real and imaginary parts of $f_1$ and $f_2$. The amplitude denoted by Rabi frequency stays constant within one piece for 20 ns with 320 pieces in total. \textbf{d,} The final waveform loaded into the AWG 81180A with four sequences combined in \textbf{c}.
}
\end{figure*}
\begin{figure*}\center
\includegraphics[width=1.0\textwidth]{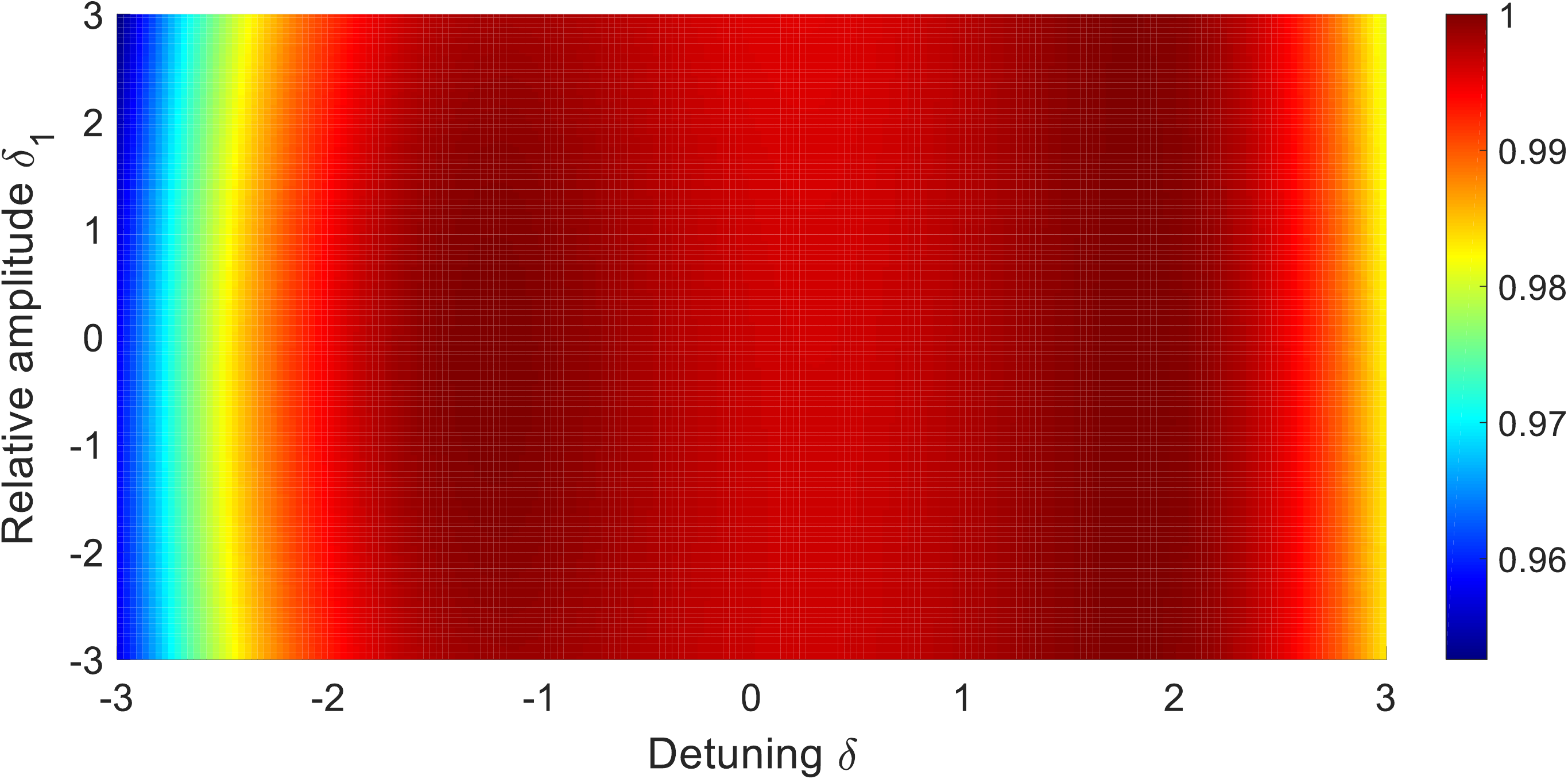}
\caption{\label{grape_fid}\textbf{Fidelity distribution for the CPhase gate.} The distribution of the gate fidelity as a function of detuning $\delta$ and relative amplitude $\delta_1$. The values of the horizontal and vertical axes are normalized relative to the standard deviations of the detuning and the relative amplitude, respectively.
}
\end{figure*}

Non-local gates play a crucial role in our experiments for both initializing nuclear spins and entanglement. They are optimized numerically by the GRAPE algorithm (gradient ascent pulse engineering). Here we illustrate the key point with the pulse sequence of the CPhase gate (Fig. \ref{grape}b). First the system Hamiltonian is reduced by abandoning the sublevel $|m_S=+1\rangle$ of the electron spin and $|m_N=-1\rangle$ of the $^{14}$N nuclear spin. It becomes
\begin{equation}
\label{Hamiltonian}
H = \omega_S S_z+\omega_N I_z^N+\omega_C I_z^C+A_\parallel S_z I_z^N +A_{zz} S_z I_z^C
\end{equation}
where $\omega_S$, $\omega_N$ and $\omega_C$ are the energy gaps of three qubits without mutual interactions, and $S_z$ and $I_z^N$ are the reduced two-level spin operators. The parameters in Eq. \ref{Hamiltonian} are all circular frequencies containing $2\pi$ in the following discussion. The transformation for the experimental interaction picture is chosen to be
\begin{equation}
\label{transformation}
U_{\rm{I}}(t) = e^{-i H t}
\end{equation}
Under this transformation, the Hamiltonian is zero with the advantage of no evolution in the waiting time of the sequence. With some reasonable rotation-wave approximations, the control Hamiltonian in the interaction picture becomes
\begin{equation}
\label{Hamiltonian_c_I}
\begin{split}
H_c^{\rm{I}}(t) = &\Omega_r^{f_1}(t) H_r^{f_1}(t)+\Omega_i^{f_1}(t) H_i^{f_1}(t)+\Omega_r^{f_2}(t) H_r^{f_2}(t)+\Omega_i^{f_2}(t) H_i^{f_2}(t)\\
&+(\Omega_N(t) I_x^N+\Omega_C(t) I_x^C)\otimes|m_S=0\rangle\langle m_S=0|\\
\end{split}
\end{equation}
with MW $f_1$ and $f_2$ applied according to Fig. \ref{grape}a and RF controlling the nuclear spins in the sublevel $|m_S=0\rangle$ of the electron spin. $\Omega_r^{f_1}(t)$, $\Omega_i^{f_1}(t)$, $\Omega_r^{f_2}(t)$, and $\Omega_i^{f_2}(t)$ are the real and imaginary parts of $f_1$ and $f_2$ optimized numerically by the GRAPE algorithm detailed in Fig. \ref{grape}c. For simplicity, RF and MW are imposed independently in our experiments. $H_r^{f_1}(t)$, $H_i^{f_1}(t)$, $H_r^{f_2}(t)$, and $H_i^{f_2}(t)$ are time-dependent, which accounts for the crosstalk between different frequencies
\begin{equation}
\label{Hamiltonian_r_f1}
\begin{split}
H_r^{f_1}(t) = &S_x\otimes P_{+1,-\frac{1}{2}}+(\cos(-A_{zz}t)S_x+\sin(-A_{zz}t)S_y)\otimes P_{+1,+\frac{1}{2}}\\
&+(\cos(-A_\parallel t)S_x+\sin(-A_\parallel t)S_y)\otimes P_{0,-\frac{1}{2}}\\
&+(\cos(-(A_\parallel+A_{zz})t)S_x+\sin(-(A_\parallel t+A_{zz})t)S_y)\otimes P_{0,+\frac{1}{2}}\\
\end{split}
\end{equation}
\begin{equation}
\label{Hamiltonian_i_f1}
\begin{split}
H_i^{f_1}(t) = &S_y\otimes P_{+1,-\frac{1}{2}}+(-\sin(-A_{zz}t)S_x+\cos(-A_{zz}t)S_y)\otimes P_{+1,+\frac{1}{2}}\\
&+(-\sin(-A_\parallel t)S_x+\cos(-A_\parallel t)S_y)\otimes P_{0,-\frac{1}{2}}\\
&+(-\sin(-(A_\parallel t+A_{zz})t)S_x+\cos(-(A_\parallel t+A_{zz})t)S_y)\otimes P_{0,+\frac{1}{2}}\\
\end{split}
\end{equation}
\begin{equation}
\label{Hamiltonian_r_f2}
\begin{split}
H_r^{f_2}(t) = &(\cos(A_{zz}t/2)S_x+\sin(A_{zz}t/2)S_y)\otimes P_{0,-\frac{1}{2}}\\
&+(\cos(-A_{zz}t/2)S_x+\sin(-A_{zz}t/2)S_y)\otimes P_{0,+\frac{1}{2}}\\
&+(\cos((A_\parallel+A_{zz}/2)t)S_x+\sin((A_\parallel+A_{zz}/2)t)S_y)\otimes P_{+1,-\frac{1}{2}}\\
&+(\cos((A_\parallel-A_{zz}/2)t)S_x+\sin((A_\parallel-A_{zz}/2)t)S_y)\otimes P_{+1,+\frac{1}{2}}\\
\end{split}
\end{equation}
\begin{equation}
\label{Hamiltonian_i_f2}
\begin{split}
H_i^{f_2}(t) = &(-\sin(A_{zz}t/2)S_x+\cos(A_{zz}t/2)S_y)\otimes P_{0,-\frac{1}{2}}\\
&+(-\sin(-A_{zz}t/2)S_x+\cos(-A_{zz}t/2)S_y)\otimes P_{0,+\frac{1}{2}}\\
&+(-\sin((A_\parallel+A_{zz}/2)t)S_x+\cos((A_\parallel+A_{zz}/2)t)S_y)\otimes P_{+1,-\frac{1}{2}}\\
&+(-\sin((A_\parallel-A_{zz}/2)t)S_x+\cos((A_\parallel-A_{zz}/2)t)S_y)\otimes P_{+1,+\frac{1}{2}}\\
\end{split}
\end{equation}
with the projection operators
\begin{eqnarray}
\label{projection}
P_{+1,-\frac{1}{2}}&=&|m_N=+1,m_C=-\frac{1}{2}\rangle\langle m_N=+1,m_C=-\frac{1}{2}|\\
P_{+1,+\frac{1}{2}}&=&|m_N=+1,m_C=+\frac{1}{2}\rangle\langle m_N=+1,m_C=+\frac{1}{2}|\\
P_{0,-\frac{1}{2}}&=&|m_N=0,m_C=-\frac{1}{2}\rangle\langle m_N=0,m_C=-\frac{1}{2}|\\
P_{0,+\frac{1}{2}}&=&|m_N=0,m_C=+\frac{1}{2}\rangle\langle m_N=0,m_C=+\frac{1}{2}|
\end{eqnarray}
In order to acquire robustness, two kinds of noise are included in numerical optimization. One is the quasi-static magnetic noise originated from local $^{13}$C bath, described by a Gaussian distribution with the standard deviation $\sigma_{\rm{mag}}=35$ kHz (calculated from the $T_2^*=6.5(3)$ \textmu s in Fig. \ref{t2star}); the other is the fluctuation of the MW amplitude, described by a Gaussian distribution with the standard deviation $\sigma_{\rm{amp}}=0.01$ relative to the mean amplitude (a coarse estimate but an overestimate). Fig. \ref{grape_fid} shows the distribution of the gate fidelity as a function of detuning $\delta$ and relative amplitude $\delta_1$. The gate fidelity is written by
\begin{equation}
\label{gate_fidelity}
F = \rm{Tr}[\it{U^\dagger(\delta,\delta_{\rm{1}})} U_{\rm{CPhase}}]/\rm{Dim}[\it{U}_{\rm{CPhase}}]
\end{equation}
where `Tr' gets the trace of a matrix and `Dim' gets the dimension of a matrix. $U_{\rm{CPhase}}$ is the ideal CPhase gate and $U(\delta,\delta_1)$ is the real gate with the noise above. The average fidelity turns out to be 0.997. It is slightly different from the fidelity of the CPhase gate in the Fig. 3d because the latter is acquired from the simulation for a full interferometer (see Section \ref{cphase_simulation}).

\subsection{Longitudinal relaxation time \textit{T}$_1$ of the NV electron spin and anomalous behavior}
\label{longitudinal_relaxation}
\begin{figure*}\center
\includegraphics[width=1.0\textwidth]{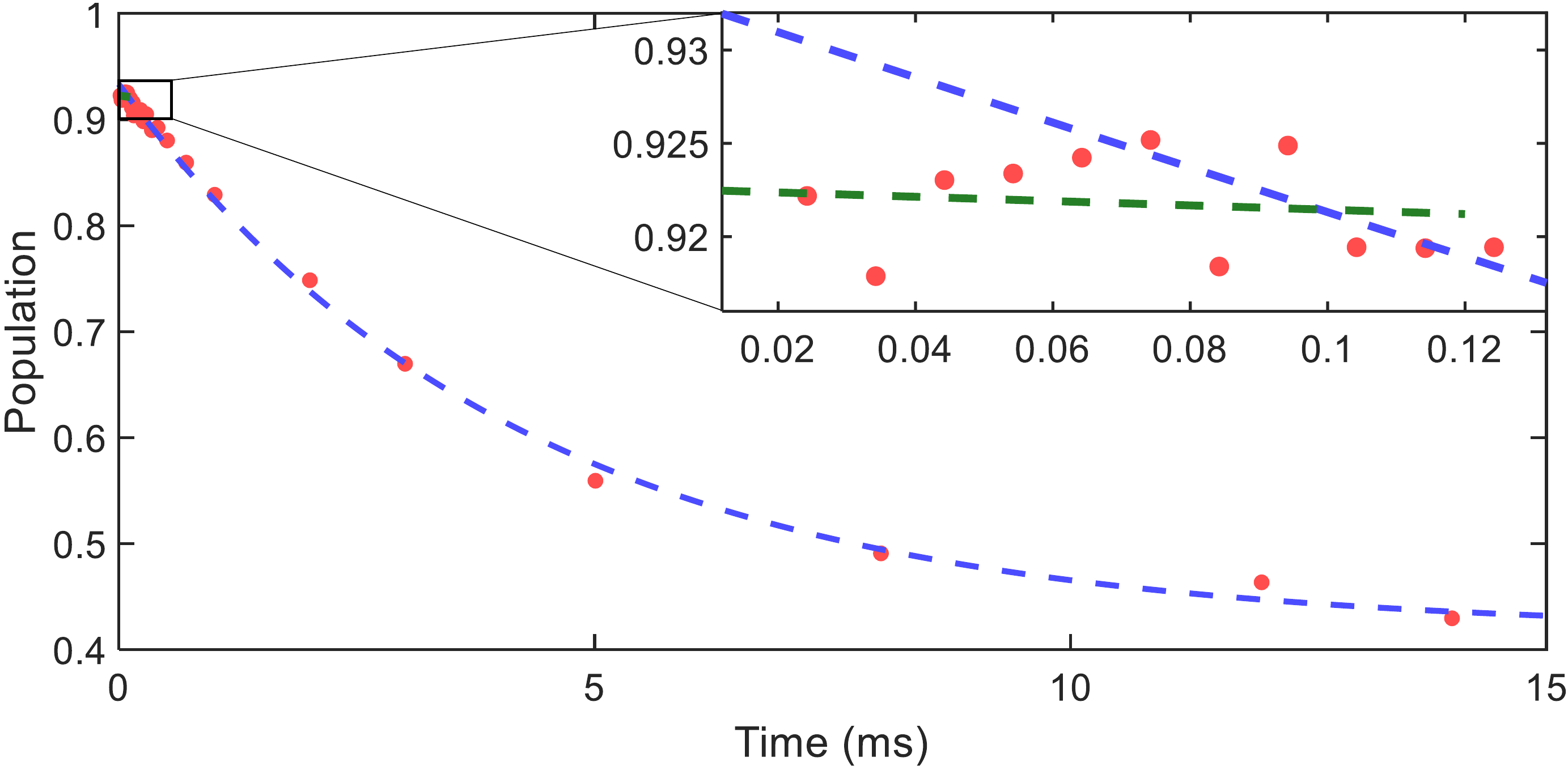}
\caption{\label{T1}\textbf{Longitudinal relaxation.} Longitudinal relaxation time $T_1$ of the NV electron spin. On the whole, the electron spin decays exponentially with the relaxation time $T_1=4.2(1)$ ms. The inset shows the relaxation behavior of the beginning 120 \textmu s. The dashed blue line is an extrapolation from the exponential decay while the green one is a linear fit. The slopes of both lines differ with a confidence beyond three standard deviations.
}
\end{figure*}
The coherence of a quantum state will be thoroughly destroyed by longitudinal relaxation characterized by the relaxation time $T_1$. The NV electron spin is featured by a long $T_1$ even at room temperature due to a high Debye temperature of diamond lattice. The relaxation time $T_1$ of our targeted NV is 4.2(1) ms obtained from Fig. \ref{T1}. However, there is an anomaly existing that the state does not relax in the beginning 120 \textmu s with a confidence beyond 3 standard deviations. To our knowledge, this anomalous and novel behavior is inexplainable and a further research is urgently needed.

The sequence time for two-spin interference is 210 \textmu s. The possibility for our quantum state to survive through 210 \textmu s is 98.5(2)$\%$ with the anomalous behavior considered. The average of the data in the beginning 120 \textmu s is 0.9218(3). Two data points 0.90485 and 0.90845 at the time of 204.3 and 234.3 \textmu s are used to estimate the survival possibility
\begin{equation}
\label{prob}
\rm{Prob_{surv}}=1-\frac{2*0.9218-0.90485-0.90845}{2} \approx 98.5\%
\end{equation}
with a coarse error estimated
\begin{equation}
\label{error}
\rm{Error}=\frac{0.90845-0.90485}{2} \approx 0.2\%
\end{equation}
For three-spin interference, the sequence is 60 \textmu s longer. The survival probability is estimated to be 97.9(3)$\%$ considering the data points 0.8978 and 0.90455 at the time of 264.3 and 294.3 \textmu s.

\pagebreak[4]

\section{Experimental setup and magnetic field stability}
\subsection{Experimental setup}
\begin{figure*}\center
\includegraphics[width=1.0\textwidth]{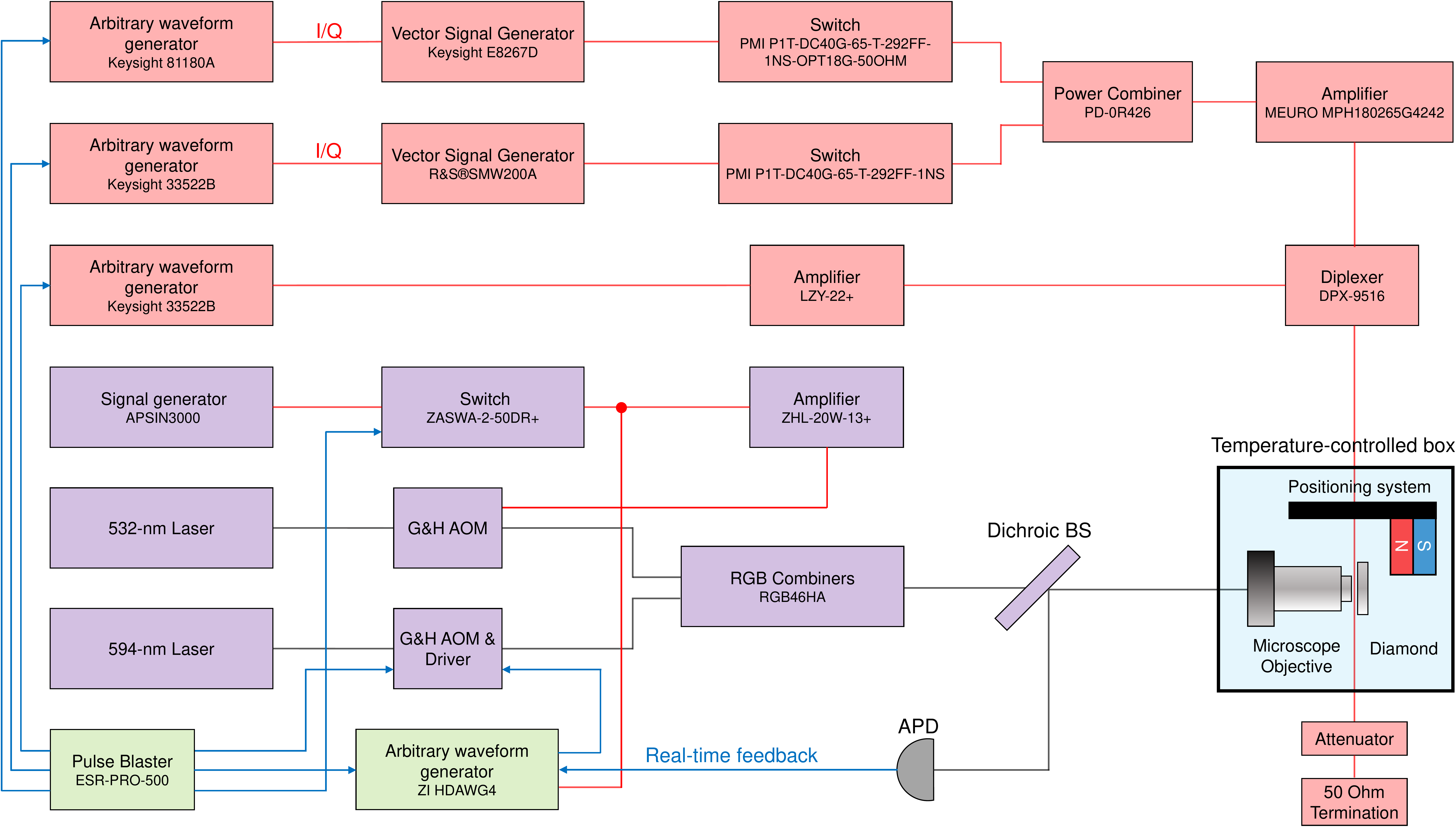}
\caption{\label{setup}\textbf{Simplified experimental setup.} The setup mainly consists of microwave (MW) and radio-frequency (RF) circuits (red), optical systems (purple), synchronization and real-time feedback system (green), and a diamond platform inside a temperature-controlled box (blue). The red, black and blue lines carry MW and RF pulses, optical beams and TTL signals respectively.
}
\end{figure*}
As shown in Fig. \ref{setup}, our experimental setup includes four parts. The first part is microwave (MW) and radio-frequency (RF) circuits. The 19.7 GHz and 25.5 GHz microwave pulses for the manipulation of the NV three sublevels are generated from two arbitrary waveform generators (AWG) with their frequencies up-converted by two vector signal generators (VSG) via I/Q modulation. And then these two paths of MW are combined, amplified, and coupled with 1-10 MHz RF for the manipulation of two nuclear spins via a diplexer. Finally, MW and RF are fed together into a coplanar waveguide microstructure for generating near-field MW and RF to influence the targeted NV. The second part is optical systems. 532-nm green laser and 594-nm orange laser for driving NV electron dynamics are emitted from two devices, while four kinds of laser beams are generated by feeding four kinds of RF with different shapes and powers into AOMs. Among them, the chopped sequence of green laser is obtained by feeding a short-pulse sequence into the green-laser AOM. Four kinds of laser beams together with sideband fluorescence (650–800 nm) go through the same oil objective, forming a confocal system. The APD collects fluorescence photons and outputs TTL signals. The third part is synchronization and real-time feedback system. The pulse blaster outputs TTL signals to synchronize all devices in our setup. The ZI AWG receives the TTL signals from the APD and performs real-time feedback for preparing NV negative state. The last part is the diamond platform located inside the temperature-controlled box. The box contains a three-dimensional positioning system holding permanent magnets, another three-dimensional positioner holding the coplanar waveguide with the diamond mounted on it, and the oil objective.

\subsection{Magnetic field stability}
\label{magnetic_stability}
\begin{figure*}\center
\includegraphics[width=1.0\textwidth]{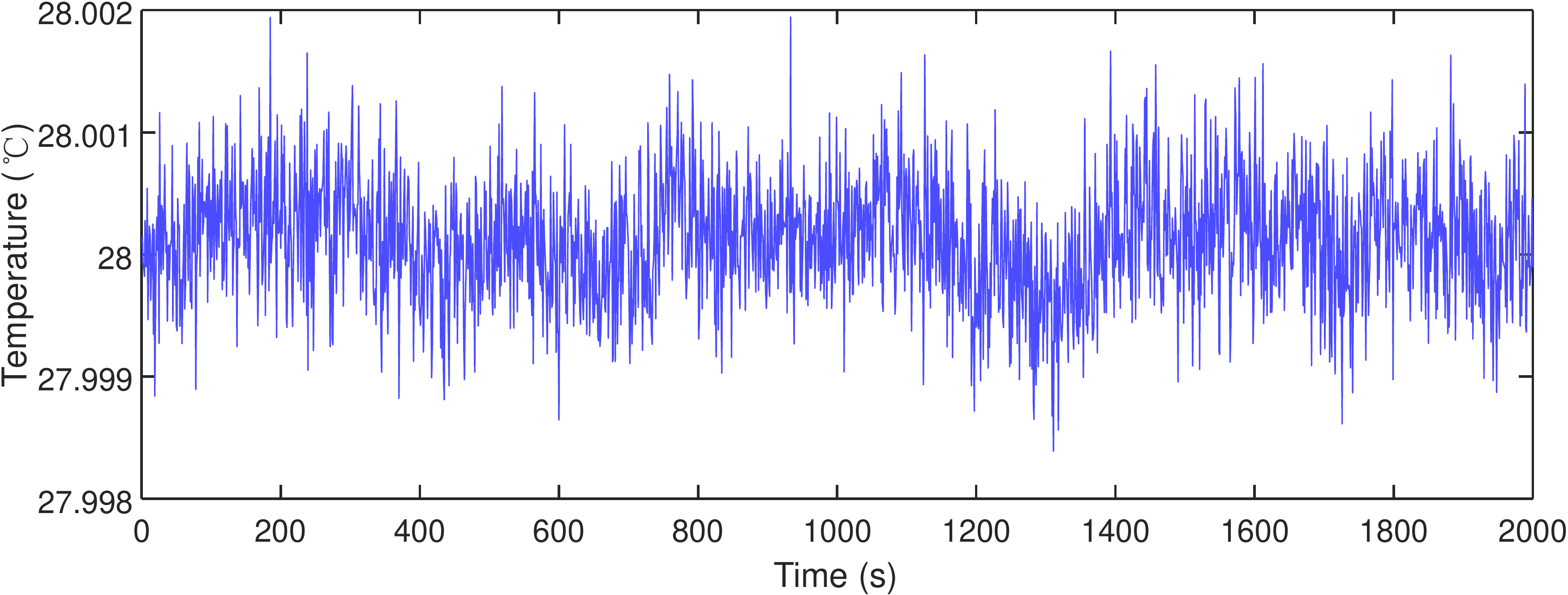}
\caption{\label{temperature}\textbf{Temperature stability.} Time trace of the temperature inside the copper box. The fluctuation is 0.5 mK (one standard deviation).
}
\end{figure*}

One of key indicators affecting the performance of our setup is magnetic field stability. It depends on many factors including temperature fluctuation, vibration and humidity fluctuation, and among them temperature fluctuation is the most important one. The temperature disturbs the magnetic field felt by the NV in two ways: first, it changes the magnetization of the magnets (0.12$\%$/\textcelsius\;for NdFeB magnets); second, it causes the stretch of materials and changes the position of the NV in the magnetic field. The latter is more remarkable especially for our high magnetic field $\sim$0.8 Tesla with the gradient $\sim$0.2 \textmu T per nanometer. With an elaborately designed configuration for heat isolation and suitable PID parameters for feedback, the temperature fluctuation is suppressed down to $\sim$0.5 mK (one standard deviation) displayed in Fig. \ref{temperature}. As for the vibration, it is well isolated by our optical table. The humidity fluctuation may become the main factor if the temperature is stable enough. In our experiment, it is observed sometimes that the fluctuation of the magnetic field is correlated with the humidity fluctuation.

\begin{figure*}\center
\includegraphics[width=1.0\textwidth]{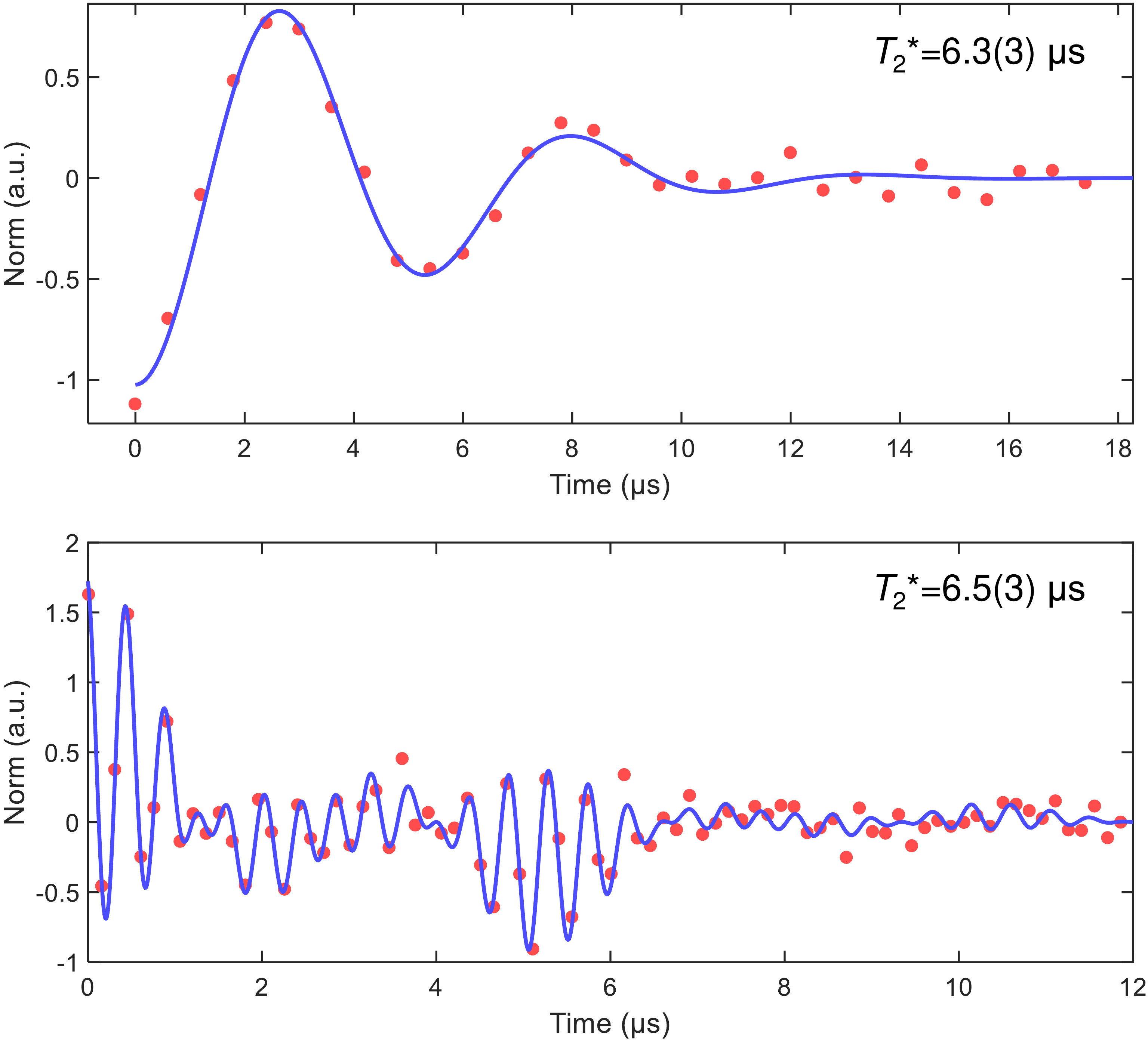}
\caption{\label{t2star}\textbf{Coherence times $T_2^*$.} Upper panel: the coherence time $T_2^*$ of Ramsey sequence under the magnetic field of 512 G. Lower panel: the coherence time $T_2^*$ under 8066 G. The $^{14}$N nuclear spin is polarized under 512 G, and no oscillation involving the $^{14}$N hyperfine interaction emerges. In comparison, the oscillation involving the $^{14}$N hyperfine interaction exists under 8066 G.
}
\end{figure*}

In order to estimate the fluctuation of the external magnetic field, the coherence times of Ramsey sequence $T_2^*$ are measured both under 512 G (the external magnetic field is stable enough and limited by the local magnetic field from nuclear spins bath) and 8066 G in Fig. \ref{t2star}. They are the same within the errors from which it is inferred that the shortening of $T_2^*$ is less than 0.5 \textmu s. Therefore, it is believed that the fluctuation is below 0.5 \textmu T (the fluctuation of the position $\sim$ 2.5 nm) during $\sim$10 minutes for data acquisition. As described in the penultimate paragraph of the text, the fluctuation is well below 1 \textmu T in two hours.
Heating from the coplanar waveguide may disturb the magnetic field felt by the NV when experimental sequences are implemented, especially RF heating due to long duration of RF pulses. Heating gives rise to the deformation of the coplanar waveguide and then changes the position of the NV. More complicated is two kinds of heating existing: one is long-term and increases the temperature gradually; the other is a short-term spike and just lasts $\sim$100 \textmu s after every RF pulse and then disappears. The latter destroys dramatically non-local gates originated from weak MW pulses. In our experiment, it is remedied by compensating every non-local gate with a frequency shift respectively. It works well for CPhase gate but not for the $\rm C_{n}NOT_{e}$ to entangle the NV electron spin (see Section \ref{local_heating}). In the future, it can be solved by designing a composite magnet with a small gradient of magnetic field.

\pagebreak[4]

\section{Deterministic initialization}
\subsection{Preparation of NV negative state}
\label{charge}
\begin{figure*}\center
\includegraphics[width=1.0\textwidth]{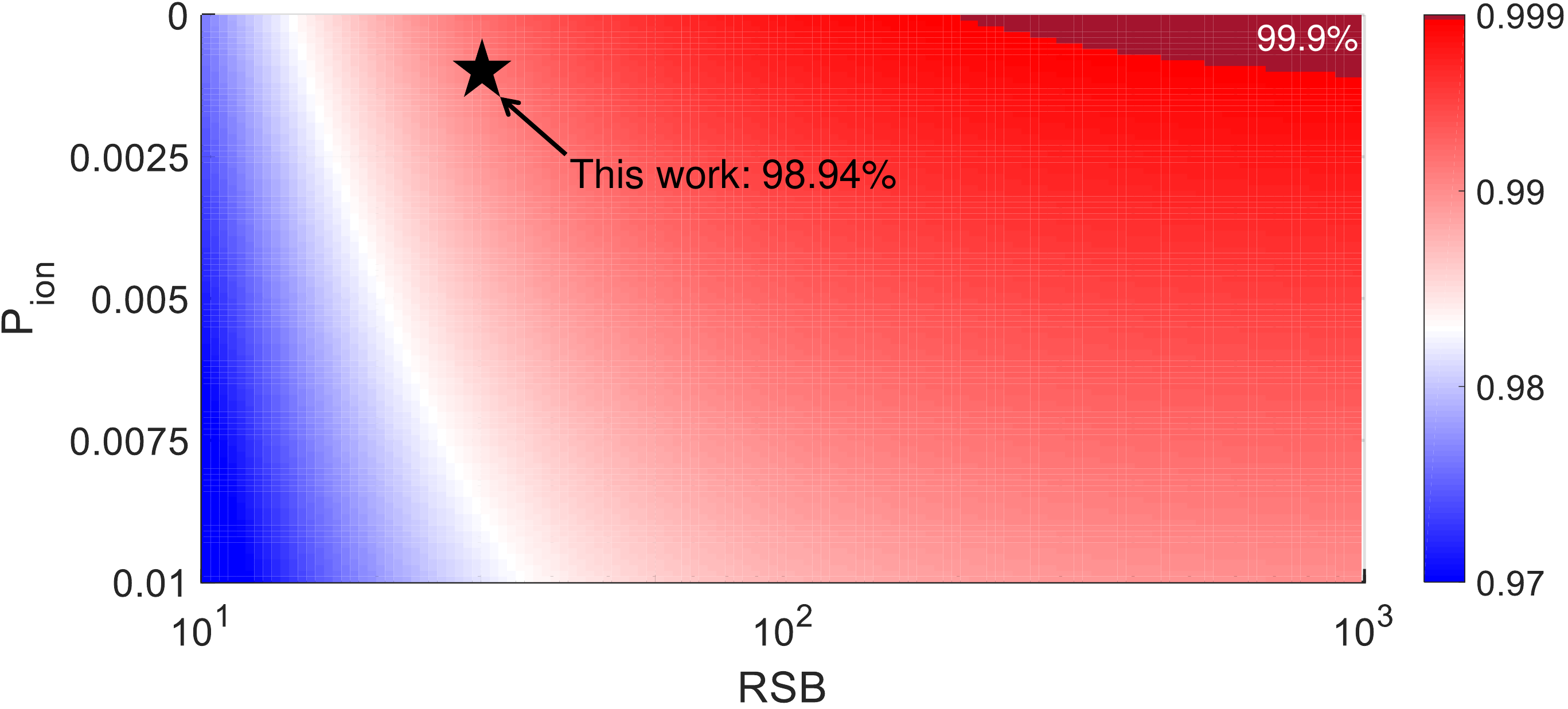}
\caption{\label{charge_error}\textbf{Fidelity distribution for NV negative state.} The distribution of the NV$^-$ fidelity as a function of RSB and $P_{\rm ion}$. It is clearly seen that the fidelity of our work is mainly limited by a low RSB.
}
\end{figure*}
The process of preparing NV negative state has been detailed in the text. Here we analyze the source of its error and expect an achievable fidelity above 0.999 in the near future. The error mainly comes from two origins: one is that the photon collected is from background fluorescence; the other is that the negative state is ionized by the 594-nm orange laser after one photon is collected. The former can be estimated by the ratio of signal fluorescence of the NV to background fluorescence (RSB). In our experiment, NV centers distribute fairly densely, and thus the fluorescence of the targeted NV contains a portion of that of a nearby NV center. The RSB is approximately 30. The latter is estimated to be $\sim$0.1$\%$ for the duration of 820 ns and the laser power 4 \textmu W in Fig. 2b of the text, the loss of the preparation fidelity by doubling the duration of the orange laser pulse. The fidelity of preparing NV negative state is
\begin{equation}
\label{NV_negative}
F \approx 1-P_{\rm{ion}}-P_{\rm{NV^0}}/(\rm{RSB}+1)
\end{equation}
where $P_{\rm{ion}}$ is the ionization probability after one photon is collected and $P_{\rm{NV^0}}$ is the probability of NV$^0$ (0.257 in Fig. 2b of the text) under the illumination of 532-nm green laser. It is plotted in Fig. \ref{charge_error}.

By the way, single-shot readout of NV charge state is performed with a weaker power $\sim$180 nW and a longer readout duration $\sim$10 ms. The outcome gives a two-peak distribution of photon number to determine which charge state it is.

\subsection{Polarization of the NV electron spin}
\label{pol_electron}
\begin{figure*}\center
\includegraphics[width=0.7\textwidth]{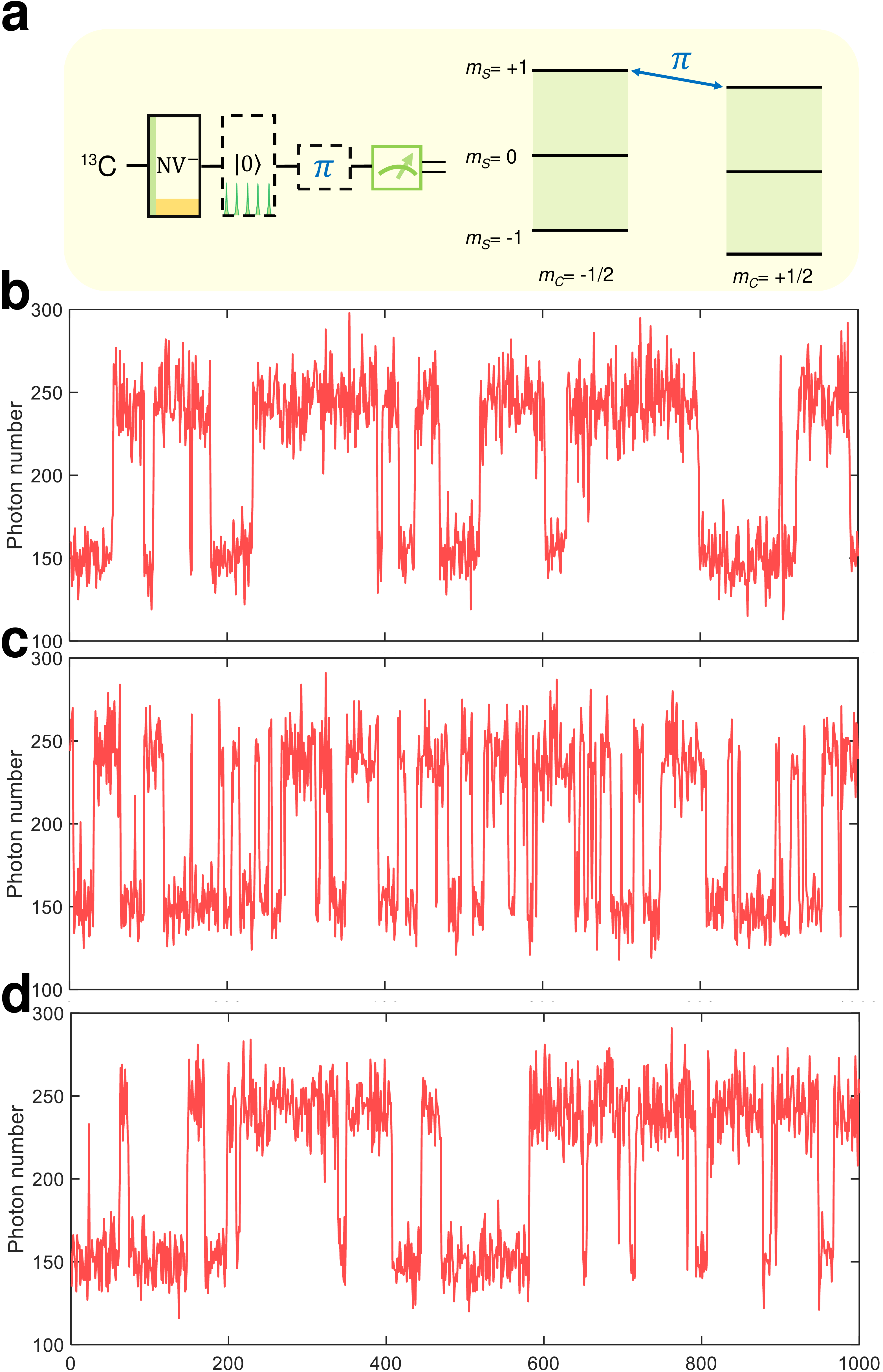}
\caption{\label{pol_electron_trace}\textbf{The measurement for the polarization of the NV electron spin.} \textbf{a,} The sequence for measuring the polarization of the electron spin. \textbf{b,} Time trace of successive projective measurements for the $^{13}$C nuclear spin without the operations denoted by dashed boxes in \textbf{a} as reference. \textbf{c,} Compared with \textbf{b}, the RF $\pi$ gate is added. \textbf{d,} Compared with \textbf{c}, the chopped laser sequence is added.
}
\end{figure*}
For NV negative state, the intersystem crossing (ISC) rate of the excited state of the triplet decaying into the singlet depends on different spin magnetic quantum numbers (Fig. 1b in the text). Such a mechanism provides a possibility for the polarization of the NV electron spin into the sublevel $|m_S=0\rangle$ and the readout of it under ambient conditions. How well the electron spin can be polarized is also limited by the branching ratio of the ground state of the singlet decaying to different magnetic sublevels of the triplet. The conventional method to polarize the electron spin is just to apply a squared pulse of 532-nm laser. However, the polarization is just at the level of $\sim$90$\%$. Because the population of the ground state of the singlet is accumulated during the illumination of green laser, and the electron spin is depolarized in accordance with the branching ratio $\sim$2-3~\cite{ISC_2016} when it decays into the ground state of the triplet. Here we employ the method of~\cite{polarization_2020} to apply a chopped laser sequence, and the population of the ground state of the singlet remains depleted all the time under the illumination of it. By applying a $\pi$ gate in the subspace of $|m_S=+1\rangle$ for $^{13}$C nuclear spin, the time trace of successive measurements of the nuclear spin is changed and the lifetimes of two states are shortened as shown in Fig. \ref{pol_electron_trace}. The residual population of the sublevel $|m_S=+1\rangle$ is estimated to be 1.13(9)$\%$ from the lifetime loss of two states. Suppose that the sublevel $|m_S=-1\rangle$ has the same population as $|m_S=+1\rangle$ (reasonable for our case), the polarization of the electron spin is 97.74(18)$\%$. As a matter of fact, such a high polarization at room temperature is not explainable based on existing models of the NV and thus the physics of the NV needs to be studied further. Besides, there is another advantage of using the chopped laser sequence that the NV negative state can survive through it with a high probability
\begin{equation}
\label{surv_prob}
\rm Prob_{surv}=98.37\%/98.94\%=99.42\%
\end{equation}

\subsection{Polarization transfer to multiple nuclear spins}
\label{pol_transfer}
\begin{figure*}\center
\includegraphics[width=1.0\textwidth]{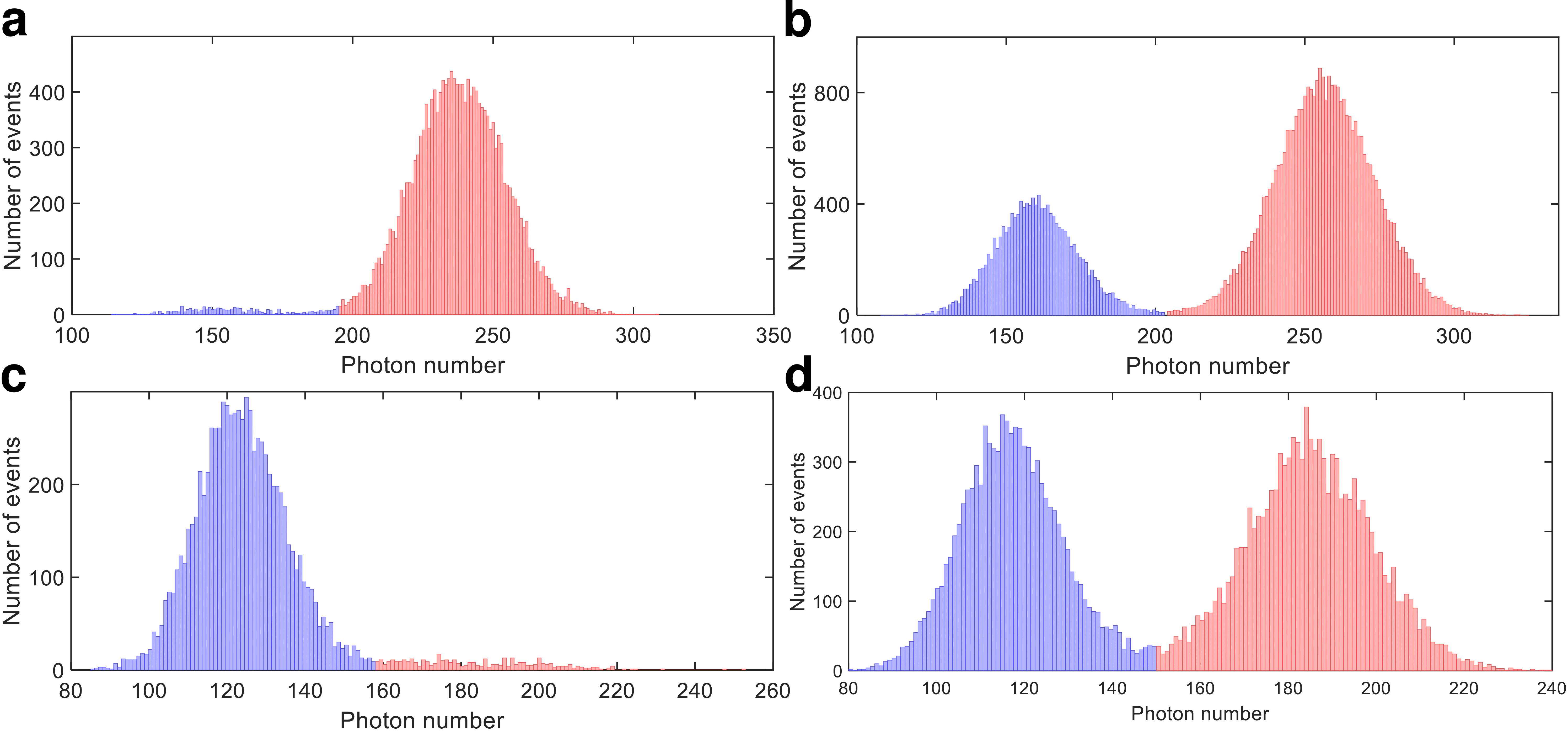}
\caption{\label{pol_transfer_histogram}\textbf{The measurement for the polarization of the nuclear spins.} \textbf{a, c,} The histograms of the projective measurements of $^{13}$C and $^{14}$N after polarizing the nuclear spins. \textbf{b, d,} The histograms of the projective measurements of $^{13}$C and $^{14}$N without polarizing the nuclear spins as reference.
}
\end{figure*}
The polarization of the electron spin is a renewable resource that can be harnessed to polarize multiple nuclear spins (here one $^{14}$N and one $^{13}$C). The maximum polarization is given by
\begin{equation}
\label{pol_nuclear}
P_{n}^{\rm{max}}=\frac{2P_{e}}{1+P_{e}}
\end{equation}
where $P_e$ is the polarization of the electron spin 97.74(18)$\%$. Hence the maximum polarization of a nuclear spin is 98.86(9)$\%$. As described in the text, the combination of optimal control (see Section \ref{local_heating}) and population shelving to the sublevel $|m_S=+1\rangle$ reduces the polarization loss to a low level. The polarization of the nuclear spins is estimated to be 98.34(13)$\%$ for $^{13}$C and 98.71(18)$\%$ for $^{14}$N by the projective measurement of each nuclear spin with the corresponding error of readout deducted (Fig. \ref{pol_transfer_histogram}). With the electron polarization removed, the left errors are 0.52(15)$\%$ for $^{13}$C and 0.15(20)$\%$ for $^{14}$N. These errors represent the loss of polarization transfer, especially evident for $^{13}$C. One possible reason is that the non-local gates for population shelving are rather coarse, especially for $^{13}$C. In the future, it can also be replaced by exquisite optimal control to arrive at the maximum polarization 98.86(9)$\%$.

\subsection{Joint initialization}
The nuclear spins survive very well under laser illumination benefitting from long lifetimes $\sim$40 ms for $^{13}$C and $\sim$10 ms for $^{14}$N. Therefore, the initialization of the nuclear spins is performed first. Before the polarization transfer, the state $|{\rm NV}^{-}, m_{S}=0\rangle$ is prepared first by real-time feedback and chopped laser sequence. After one round of polarization transfer, the electron spin is polarized again without preparing NV$^-$ because the chopped laser sequence just destroys NV$^-$ by a small error $0.57(5)\%$. With three rounds of polarization transfer, both $^{13}$C and $^{14}$N nuclear spins are polarized but the polarization is mainly limited by the fidelity of NV$^-$. The process above is implemented 2-3 times to cancel the impact of imperfect NV$^-$. After the initialization of two nuclear spins, the electron state $|{\rm NV}^{-}, m_{S}=0\rangle$ is prepared again. The joint state $|{\rm NV}^{-}, m_{S}=0\rangle \otimes |m_{C}=-1/2\rangle \otimes |m_{N}=+1\rangle$ is initialized deterministically with a fidelity of 93.3(3)$\%$.

\pagebreak[4]

\section{Error analysis}
\subsection{The error of the CPhase gate by simulation}
\label{cphase_simulation}
The CPhase gate is well done with a high fidelity above fault-tolerant threshold. The residual error is too small to estimate by process tomography. Here we estimate the error through a numerical simulation for our two-spin interference. With a phase $\phi$ accumulated, the fidelity is calculated as
\begin{equation}
\label{fidelity_interference}
F_\phi = \frac{1}{N}\sum_{i=1}^{N} |\langle\psi_{\rm{id}}(\phi)|\psi_{\delta^i,\delta_1^i}(\phi)\rangle|^2
\end{equation}
with the ideal state $|\psi_{\rm{id}}(\phi)\rangle$ and the real state $|\psi_{\delta^i,\delta_1^i}(\phi)\rangle$ with two kinds of noise (see Section \ref{local_heating})
\begin{eqnarray}
\label{ideal_state}
|\psi_{\rm{id}}(\phi)\rangle &=& e^{-i\frac{\pi}{2} I_x^C}U_{\rm{CPhase}}e^{-i\frac{\pi}{2} I_y^N}e^{-i\phi(I_z^C+I_z^N)}e^{-i\frac{\pi}{2} I_y^C}U_{\rm{CPhase}}e^{i\frac{\pi}{2} I_y^C}e^{-i\frac{\pi}{2} I_y^N}|\psi_0\rangle\\
|\psi_{\delta^i,\delta_1^i}(\phi)\rangle &=& e^{-i\frac{\pi}{2} I_x^C}U(\delta^i,\delta_1^i)e^{-i\frac{\pi}{2} I_y^N}e^{-i\phi(I_z^C+I_z^N)}e^{-i\frac{\pi}{2} I_y^C}U(\delta^i,\delta_1^i)e^{i\frac{\pi}{2} I_y^C}e^{-i\frac{\pi}{2} I_y^N}|\psi_0\rangle
\end{eqnarray}
with the initial state $|\psi_0\rangle=|m_{S}=0\rangle \otimes |m_{C}=-1/2\rangle \otimes |m_{N}=+1\rangle$. By varying different phases $\phi$ from $-\pi$ to $\pi$ with a step of $\pi/4$, the average fidelity is given by 99.5(2)$\%$ for the error analysis in the next subsection. In the future, the fidelity can be measured experimentally by adopting the method of~\cite{benchmarking_2015}.

\subsection{Error analysis for multi-qubit entangled interference}
\label{error_analysis}
\begin{figure*}\center
\includegraphics[width=1.0\textwidth]{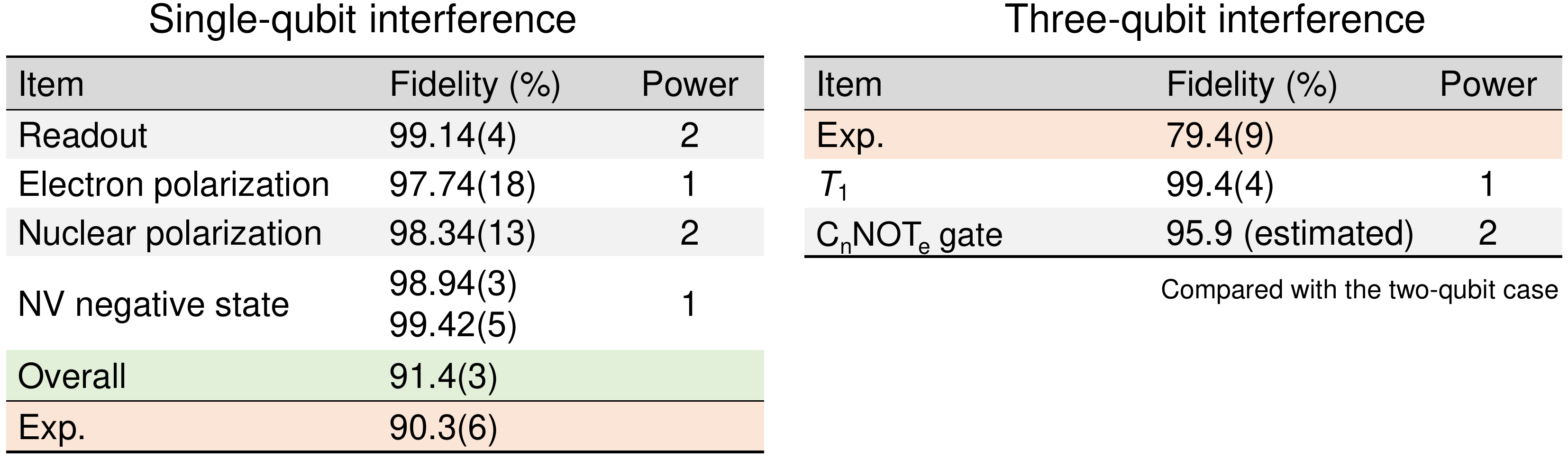}
\caption{\label{error}\textbf{Error analysis for one-spin and three-spin interference.} Left panel: error budget for one-spin interference; right panel: the estimated fidelity of the $\rm C_{n}NOT_{e}$ gate conditional on the $^{13}$C nuclear spin.
}
\end{figure*}
The sections above have carefully studied all kinds of experimental imperfections, including readout (in the text), electron polarization (Section \ref{pol_electron}), nuclear polarization (Section \ref{pol_transfer}), NV negative state (Section \ref{charge}), CPhase gate (Section \ref{cphase_simulation}) and $T_1$ (Section \ref{longitudinal_relaxation}) in Fig. 3d of the text. The interference visibility is given by the overall fidelity
\begin{equation}
\rm{Vis}=\it{F}=\prod_i \rm Fidelity_i^{Power_i}
\end{equation}
`Vis' is the same as that in Section \ref{QFI_exp}. Fig. \ref{error} shows the error analysis for one-spin and three-spin interference. The three-spin interference has an extra error from two $\rm C_{n}NOT_{e}$ gates optimized by the GRAPE algorithm as well. The fidelity is supposed to be at least 0.99. However, compared with the two-spin case, the visibility of three-spin interference decreases by 7.5$\%$. It is likely that short-term heating causes the electron spin to decohere because the superposition of the electron spin is very susceptible to the fluctuation of the magnetic field (see Section \ref{magnetic_stability}). Inferred from the visibility with all other errors deducted, the fidelity of the $\rm C_{n}NOT_{e}$ gate conditional on the $^{13}$C nuclear spin is approximately 95.9$\%$.

\subsection{Expected scaling of quantum Fisher information with spin number}
With the experimental errors analyzed above, a scaling of quantum Fisher information with spin number is expected. Adding an additional nuclear spin incurs additional errors $\sim$4$\%$ including imperfect polarization $\sim$1.5$\%$ (the average of $^{13}$C and $^{14}$N) and additional error of non-local gates $\sim$1$\%$. Starting from the visibility of one-spin interference $\sim$0.91, the quantum Fisher information (QFI) for $N$-spin interference is expected to be
\begin{equation}
\label{scaling}
\rm{QFI}(\it{N}) \approx N^{\rm{2}}(\rm{0.91\times 0.96}^{\it{(N-1)}})^{\rm{2}}
\end{equation}
It saturates at 24 spins with the maximum 67.

\pagebreak[4]

\bibliographystyle{Nature}

\clearpage